\documentclass[lettersize,10pt,fleqn]{article}
\usepackage{graphicx}
\usepackage{subfigure}
\usepackage{morefloats}
\usepackage{color}
\usepackage{setspace}
\usepackage{floatrow}
\usepackage{bm}
\usepackage{color}
\usepackage{float}
\usepackage{amssymb}
\usepackage{geometry}%
\usepackage{amsmath}
\usepackage{CJK} 
\usepackage[colorlinks,linkcolor=blue,citecolor=blue,bookmarks,pdfstartview=FitH]{hyperref}
\usepackage{amsthm,amsmath,amssymb}
\usepackage{mathrsfs}
\usepackage{authblk}
\usepackage{cancel}  
\usepackage{makecell}
\usepackage{comment}
\usepackage[bottom]{footmisc}
\begin{document}

\title{\textbf{MAS: A versatile Landau-fluid eigenvalue {\color{black}code} for plasma stability analysis in general geometry}}

\author[1,2,\thanks{jbao@iphy.ac.cn}]{J. Bao}
\author[1,2,3,\thanks{Author to whom correspondence should be addressed: wzhang@iphy.ac.cn}]{W. L. Zhang}
\author[1,2,3]{D. Li}
\author[4]{Z. Lin}
\author[5]{G. Dong}
\author[6]{C. Liu}
\author[7,8]{H. S. Xie}
\author[9]{G. Meng}
\author[10]{J. Y. Cheng}
\author[1,2]{C. Dong}
\author[1,2]{J. T. Cao}
\date{} 
\affil[1]{\small Beijing National Laboratory for Condensed Matter Physics and CAS Key Laboratory of Soft Matter Physics, Institute of Physics, Chinese Academy of Sciences, Beijing 100190, China}
\affil[2]{\small University of Chinese Academy of Sciences, Beijing 100049, China}
\affil[3]{\small Songshan Lake Materials Laboratory, Dongguan, Guangdong 523808, China}
\affil[4]{\small University of California, Irvine, California, 92697, USA}
\affil[5]{\small Energy Singularity Co., Ltd., Shanghai, China}
\affil[6]{\small Princeton Plasma Physics Laboratory, Princeton University, PO Box 451, Princeton, New Jersey 08543, USA}
\affil[7]{\small Hebei key laboratory of compact fusion, Langfang, 065001, China}
\affil[8]{\small ENN Science and Technology Development Co., Ltd., Langfang 065001, China}
\affil[9]{\small Max Planck Institut für Plasmaphysik, 85748 Garching, Germany}
\affil[10]{\small Department of Physics, University of Colorado at Boulder, Boulder, Colorado 80309, USA}
\maketitle
\begin{abstract}
We have developed a new global eigenvalue code, Multiscale Analysis for plasma Stabilities (MAS), for studying plasma problems with wave toroidal mode number ($n$) and frequency ($\omega$) in a broad range of interest in general tokamak geometry, based on a five-field Landau-fluid description of thermal plasmas. Beyond keeping the necessary plasma fluid response, we further retain the important kinetic effects including diamagnetic drift, ion finite Larmor radius, finite parallel electric field ($E_{||}$), ion and electron Landau resonances in a self-consistent and non-perturbative manner without sacrificing the attractive efficiency in computation. The physical capabilities of the code are evaluated and examined in the aspects of both theory and simulation. In theory, the comprehensive Landau-fluid model implemented in MAS can be reduced to the well-known ideal MHD model, electrostatic ion-fluid model, and drift-kinetic model in various limits, which clearly delineates the physics validity regime. In simulation, MAS has been well benchmarked with theory and other gyrokinetic and kinetic-MHD hybrid codes in a manner of adopting the unified physical and numerical framework, which covers the kinetic Alfv\'en wave (KAW), ion sound wave (ISW), low-$n$ kink, high-$n$ ion temperature gradient mode (ITG) and kinetic ballooning mode (KBM).  Moreover, MAS is successfully applied to model the Alfv\'en eigenmode (AE) activities in DIII-D discharge \#159243, which faithfully captures the frequency sweeping of reversed shear Alfv\'en eigenmode (RSAE), the tunneling damping of toroidal Alfv\'en eigenmode (TAE), as well as the polarization characteristics of kinetic beta-induced Alfv\'en eigenmode (KBAE) and beta-induced Alfv\'en-acoustic eigenmode (BAAE) being consistent with former gyrokinetic theory and simulation. With respect to the key progress contributed to the community, MAS has the advantage of combining rich physics ingredients, realistic global geometry and high computation efficiency together for plasma stability analysis in linear regime.
\end{abstract}

\section{Introduction}
With the injection of high power radio frequency waves and neutral-beam injection for heating and current drive, the plasma confinement properties of fusion reactors are mostly determined by various types of instabilities as the consequences of releasing the excessive free energy in the distribution functions, which give rise to the plasma profile relaxations through disruption events \cite{hender2007}, anomalous transports \cite{doyle2007,lin1998}, {\color{black}energy particle (EP)} losses \cite{fasoli2007,heidbrink2008, heidbrink2020} etc. Specifically, the global Magnetohydrodynamic(MHD) activities are the most crucial and fundamental constraints for limiting the plasma current and pressure values in fusion devices \cite{troyon1984}, which can be violent such as sawtooth crash \cite{kadomtsev1975} and neoclassical tearing mode \cite{qu1985}, and generally destroy the overall equilibrium stability and terminate the plasma confinement. In contrast to the macroscopic MHD instabilities, the microscopic drift-wave instabilities are mild and lead to the plasma turbulent (anomalous) transport, which are mostly driven by pressure gradients of each plasma species in the ambient nonuniform magnetic field. Great efforts have been made for understanding the mechanisms of excitation, saturation and turbulent transport for various types of drift-wave instabilities, which restrict the confinements of particle, momentum and energy in nowadays tokamaks \cite{horton1999}. Moreover, the population of EPs can be increased through external auxiliary heating and internal fusion reaction (i.e., $\alpha$-particle) processes, consequently EPs can drive originally stable Alfv\'en eigenmodes (AEs) unstable \cite{fu89,cheng92}, and generate new instabilities such as fishbone \cite{chen1984} and EP modes (EPMs) \cite{zonca2000} owing to resonant wave-particle interactions. Due to the large finite Larmor radius (FLR) and finite orbit width (FOW) of EPs, these EP-driven instabilities are usually characterized by mesoscale or macroscale electromagnetic perturbations, which in turn induce large EP transports and lead to the fusion power loss. In addition, nowadays tokamak geometry is no longer the concentric circular, the D shape, triangularity and elongation effects play important roles to increase the device $\beta$ limit (where $\beta = 8\pi P_0/B_0^2$ is the ratio of the plasma pressure to the magnetic pressure). Given considerations to above factors, comprehensive modelings of macroscale MHD modes, mesoscale EP-driven instabilities such as AEs and microscale drift-wave instabilities using realistic experimental geometry, are necessary for understanding the experimental phenomena, predicting the overall plasma confinement, and optimizing the fusion power plant design \cite{fasoli2016}.

There are two important approaches for modeling fusion plasma, which deal with the time derivative operator differently, i.e., eigenvalue approach by applying Fourier transformation $\partial_t\to-i\omega$ and initial value approach by discretizing the operator numerically $\partial_t\to1/\Delta t$. There is no doubt that the initial value approach is more powerful in addressing complex nonlinear physics phenomenon since it allows the nonlinear dynamics such as mode-mode coupling/beating. But, eigenvalue approach is more efficient and accurate than the initial value approach in the linear regime, which does not evolve the dynamics of physics quantities and suffer the numerical error of time domain discretization. {\color{black}Historically, many eigenvalue  codes have been developed based on the ideal or resistive MHD model for the toroidal plasmas, such as NOVA \cite{nova1987}, CASTOR \cite{kerner1998} and MARS \cite{bondeson1994} for low-$n$ kinks, gap-mode type AEs and resistive wall modes (RWMs) , and ELITE \cite{elite2002} for high-$n$ ballooning modes. The corresponding kinetic-MHD hybrid versions, NOVA-K \cite{cheng92}, CASTOR-K \cite{borba1999} and MARS-K \cite{liu2008}, can incorporate the kinetic effects using an extended energy principle such as wave-particle resonance, FLR and FOW etc., and recalculate the stability spectra either in a perturbative \cite{cheng92,borba1999} or a nonperturbative manner \cite{liu2008}. However, most kinetic-MHD hybrid eigenvalue codes use one-fluid MHD model for bulk plasmas by ignoring its kinetic effects, which is good for describing high frequency Alfv\'enic spectra that are mostly fluid type. It is later found theoretically that the kinetic thermal ion (KTI) compression beyond the MHD physics can dramatically alter the polarization of low frequency AEs such as for the {\color{black}beta-induced Alfv\'en eigenmode (BAE)} \cite{zonca1996,chavdarovski2009}, where acoustic spectra and Alfv\'enic spectra couple together and should be treated on the same footing by using physics model with essential kinetic effects. Regarding to the fully gyrokinetic simulation using eigenvalue approach, LIGKA \cite{lauber2007} is developed for solving the gyrokinetic moment (GKM) and quasi-neutrality (QN) equations, which can compute both the acoustic and Alfv\'enic spectra using the exact distribution function in 5D phase space and has the advantage in low frequency AE analysis \cite{lauber2008,lauber2009}. {\color{black}Besides the gyrokinetic model, it is worth mentioning that, the Landau closure technique for fluid approach (termed as Landau-fluid) can capture the crucial kinetic effects such as Landau resonance only using 3D spatial coordinates, and the closure operator is first formulated in Fourier space for physical problems with small spatial inhomogeneities \cite{hammett1990} and later extended to the configuration space for better incorporating the plasma nonuniformity, boundary and geometry effects \cite{dimits2014}, which greatly saves the computing cost by avoiding velocity domain calculation and has been widely used for plasma turbulence simulations \cite{synder2000,ma2016}.} Recently, FAR3D \cite{varela2017} captures the EP-AE Landau resonance in the well-circulating limit (i.e., assume zero pitch angle $\lambda = \mu B_a/E = 0$ for all particles) by using a modified Landau closure \cite{spong2013}, which is numerically efficient for computing the unstable AE spectra in 2D and 3D experimental geometries.} {\color{black}It should also be noted that above eigenvalue codes apply Fourier expansion on the perturbations along poloidal and toroidal angles, which requires to keep a large number of poloidal harmonics in the high-$n$ regime for toroidal coupling. For comparison, the well-known ballooning representation is well-established for short wave-length modes that satisfy the scale-separation from equilibrium \cite{conner1979}, which transforms the two-dimensional physical problem in the poloidal plane to two one-dimensional problems (i.e., solving the parallel mode structure in the lowest order and the radial envelope in the higher order), thus the ballooning approach is more efficient at incorporating the toroidal coupling effect for both theory and simulation when the radial envelope variation can be ignored in the high-$n$ regime. For example, the theoretical framework of general fishbone-like dispersion relation (GFLDR) \cite{zonca2014} applies the ballooning approach and has many practical applications on EP physics, and DAEPS code \cite{li2020} is developed based on GFLDR that keeps the properly asymptotic behavior in ballooning space. Meanwhile, the fluctuation polarization information can be clearly reflected by the eigenvector solutions in ballooning space, for example, FALCON code \cite{falcon2019} computes the MHD continuous spectra based on ballooning approach and reveals the mixed polarization of shear Alfv\'en wave (SAW) and ion sound wave (ISW) through the Alfv\'enicity parameter (i.e., the degree of Alfv\'enic polarization) on dispersion curves \cite{falcon2020}.}

In fusion plasmas, each particle species contributes to both the damping and driving of plasma spectra, and the relative importance depends on the specific wave-particle resonance and plasma gradients. Thus, in order to make reliable predictions for various plasma stabilities, both the stabilizing and destabilizing effects of each species should be kept in the physics model for properly calculating the damping and growth rates. However, many plasma stability codes based on kinetic-MHD and Landau-fluid models are lack of adequate kinetic effects of bulk plasmas associated to damping and driving. The nonperturbative gyrokinetic approach \cite{lauber2007} can deal with the kinetic effects, but requires a much increased computational cost compared to the perturbative kinetic-MHD approach \cite{cheng92}. To strike the balance between these two methodologies in opposite limits, the nonperturbative Landau-fluid approach \cite{hammett1990} adopts the particular closures in order to guarantee the response function being close to the gyrokinetic model, which can compute the plasma spectra efficiently with certain kinetic effects in the linear regime. 

The purpose of this work is to demonstrate a new plasma stability eigenvalue code developed in the global and experimental geometry, namely, MAS with a five-field $\left(\delta\phi,\delta A_{||},\delta P_i, \delta u_{||i}, \delta n_i\right)$ Landau-fluid physics model formulated for bulk thermal plasmas, which incorporates the essential kinetic effects associated to both driving and damping processes. MAS physics model has three key characteristics: First the fully electromagnetic perturbations are considered, where the electrostatic potential $\delta\phi$ and parallel vector potential $\delta A_{||}$ are explicitly evolved and the leading $\delta B_{||}$ effect of drift reversal cancellation is implicitly imposed. Second the thermal ion dynamics is treated using an extended drift-ordering, which keeps the ion-diamagnetic drift, ion-FLR and ion-Landau damping effects. Third the thermal electron physics not only covers the adiabatic electron response widely used in the gyrokinetic simulation model \cite{holod2009, bao2017} (including the finite parallel electric field $E_{||}$ and electron-diamagnetic drift effects in the lowest order), but also further extends with the electron-Landau damping on the side of adiabatic regime ($\omega < k_{||}v_{the}$). We then show our physics model can reduce to the well-known ideal reduced/full MHD, electrostatic ion-fluid and drift-kinetic models in various physics limits, which constitute a hierarchy of models from simple to complex as a theoretical benchmark. After implementing the physics model into a general tokamak geometry represented in Boozer coordinates, we present MAS multi-scale physics capabilities of low-$n$ MHD modes, mediate-$n$ AEs and high-$n$ drift wave instabilities by carrying out a series of benchmarks, which show good agreements with other fusion codes. Moreover, It is worthwhile mentioning that MAS can resolve MHD singularities such as the Alfv\'enic and acoustic continuum resonances in a physical manner by retaining the small-scale physics terms in the higher order, rather than using the extra numerical viscosity in most kinetic-MHD hybrid codes. In addition, MAS has a few early benchmarks and applications on EAST and DIII-D experiments with simplified models during code development \cite{bao2020,zhao2021,brochard2022}, which are covered by the comprehensive Landau-fluid model presented in this work. 

This paper is organized as follows. The theoretical formulation of electromagnetic Landau-fluid physics model in MAS is introduced in section \ref{2}. In section \ref{numerical_method}, we present the numerical schemes on coordinate system, operator discretization and matrix construction for solving the physics equation set in general geometry with complexity. The reductions of Landau-fluid model in various limits as theoretical benchmarks are given in section \ref{3}, and the numerical benchmarks of MAS on waves and instabilities of reactive- and dissipative-type are shown in section \ref{4}, in both low- and high-$n$ regimes. The summary is given in section \ref{5}. The orientation convention of flux coordinate system in MAS is introduced in appendix \ref{A5}. Appendices \ref{A3} and \ref{A4} provide additional supplements on ideal full-MHD and drift-kinetic models, which are relevant to MAS theoretical benchmarks.

\section{Electromagnetic Landau-fluid physics model}\label{2}
An electromagnetic Landau-fluid model is applied in MAS code, which consists of dynamic equations for $\left(\delta\phi, \delta A_{||}, \delta P_i, \delta u_{||i}, \delta n_i\right)$ based on the {\color{black} extended drift-ordering, i.e., besides the resistive MHD physics, the present Landau-fluid model faithfully preserves kinetic effects including ion and electron diamagnetic drifts, ion FLR, ion and electron Landau damping effects in a self-consistent and non-perturbative manner.} Compared to many other tokamak stability codes using one-fluid ideal/resistive MHD description of thermal plasmas, MAS Landau-fluid model covers a broad frequency range from bulk plasma diamagnetic frequency to Alfv\'en frequency and a wide spatial range from ion Larmor radius to machine minor radius. The governing equations of the Landau-fluid model include \\
the vorticity equation
\begin{flalign}\label{vorticity}
\begin{split}
&\frac{\partial }{\partial t}\frac{c}{V_A^2}\nabla_\perp^2\delta\phi
\underbrace{+
\frac{\partial }{\partial t}\left(0.75\rho_i^2\nabla_\perp^2\right)\frac{c}{V_A^2}\nabla_\perp^2\delta\phi}_{\{Ion-FLR\}}
\underbrace{+
i\omega_{*p,i}\frac{c}{V_A^2}\nabla_\perp^2\delta\phi}_{\{Drift\}}
+\mathbf{B_0}\cdot\nabla\left(\frac{1}{B_0}\nabla_\perp^2\delta A_{||}\right) 
-\frac{4\pi}{c}\boldsymbol{\delta B}\cdot\nabla\left(\frac{J_{||0}}{B_0}\right)\\
&- 8\pi\left(\nabla\delta P_i + \nabla\delta P_e\right)\cdot\frac{\mathbf{b_0}\times\boldsymbol{\kappa}}{B_0}
=0,
\end{split}
\end{flalign}
the parallel Ohm's law
\begin{flalign}\label{ohm_law}
\begin{split}
\frac{\partial \delta A_{||}}{\partial t} =
& -c\mathbf{b_0}\cdot\nabla\delta \phi 
\underbrace{+ \frac{cT_{e0}}{en_{e0}}\mathbf{b_0}\cdot\nabla\delta n_e
+ \frac{cT_{e0}}{en_{e0}B_0}\boldsymbol{\delta B}\cdot\nabla n_{e0}}_{\{Drift\}}
\underbrace{+\frac{cm_e}{e}\sqrt{\frac{\pi}{2}}v_{the}|k_{||}|\delta u_{||e}}_{\{Electron-Landau\}}
\underbrace{+ \frac{c^2}{4\pi}\eta_{||}\nabla_\perp^2\delta A_{||}}_{\{Resistivity\}}
\end{split},
\end{flalign}
the ion pressure equation
\begin{flalign}\label{ion_pressure}
\begin{split}
\frac{\partial \delta P_i}{\partial t} 
&+ \frac{c\mathbf{b_0}\times\nabla\delta\phi}{B_0}\cdot\nabla P_{i0} 
+2\Gamma_{i\perp} P_{i0}c\nabla\delta\phi\cdot\frac{\mathbf{b_0}\times\boldsymbol{\kappa}}{B_0}
+\Gamma_{i||} P_{i0}\mathbf{B_0}\cdot\nabla\left(\frac{\delta u_{||i}}{B_0}\right)
\underbrace{-i\Gamma_{i\perp} \omega_{*p,i}Z_in_{i0}\rho_i^2\nabla_\perp^2\delta \phi}_{\{Ion-FLR\}}\\
&\underbrace{+2\Gamma_{i\perp} P_{i0}\frac{c}{Z_i}\nabla\delta T_i\cdot\frac{\mathbf{b_0}\times\boldsymbol{\kappa}}{B_0}
+2\Gamma_{i\perp} T_{i0}\frac{c}{Z_i}\nabla\delta P_i\cdot\frac{\mathbf{b_0}\times\boldsymbol{\kappa}}{B_0}}_{\{Drift\}}
\underbrace{+n_{i0}\frac{2}{\sqrt{\pi}}\sqrt{2}v_{thi}|k_{||}|\delta T_i}_{\{Ion-Landau\}}
=0
\end{split},
\end{flalign}
the parallel momentum equation
\begin{flalign}\label{sound}
\begin{split}
m_in_{i0}\frac{\partial \delta u_{||i}}{\partial t} 
=& -\mathbf{b_0}\cdot\nabla\delta P_e
-\frac{1}{B_0}\boldsymbol{\delta B}\cdot\nabla P_{e0}
-\mathbf{b_0}\cdot\nabla\delta P_i
-\frac{1}{B_0}\boldsymbol{\delta B}\cdot\nabla P_{i0}\\
&\underbrace{- Z_in_{i0}\frac{m_e}{e}\sqrt{\frac{\pi}{2}}v_{the}|k_{||}|\delta u_{e||}}_{\{Electron-Landau\}} 
\underbrace{- Z_in_{i0}\eta_{||}\frac{c}{4\pi}\nabla_\perp^2\delta A_{||}}_{\{Resistivity\}}
\end{split},
\end{flalign}
and the ion continuity equation
\begin{flalign}\label{ion_density}
\begin{split}
\frac{\partial \delta n_i}{\partial t} 
+ \frac{c\mathbf{b_0}\times\nabla\delta\phi}{B_0}\cdot\nabla n_{i0} 
+ 2cn_{i0}\nabla\delta\phi \cdot\frac{\mathbf{b_0}\times\boldsymbol{\kappa}}{B_0}
+ n_{i0}\mathbf{B_0}\cdot\nabla\left(\frac{\delta u_{||i}}{B_0}\right)
\underbrace{- i \omega_{*p,i}\frac{Z_in_{i0}}{T_{i0}}\rho_i^2\nabla_\perp^2\delta\phi}_{\{Ion-FLR\}}
\underbrace{+ \frac{2c}{Z_i}\nabla\delta P_i\cdot\frac{\mathbf{b_0}\times\boldsymbol{\kappa}}{B_0}}_{\{Drift\}}
=0
\end{split},
\end{flalign}
where $J_{||0} = \frac{c}{4\pi}\mathbf{b_0}\cdot\nabla\times\mathbf{B_0}$ is the parallel equilibrium current density, $\boldsymbol{\kappa} = \mathbf{b_0}\cdot\nabla \mathbf{b_0}$ is the magnetic field curvature, $\rho_i = v_{thi}/\Omega_{ci}$ is the ion Larmor radius, $v_{thi} = \sqrt{T_{i0}/m_i}$, $\Omega_{ci} = Z_iB_0/\left(cm_i\right)$, $\omega_{*p,i} =\omega_{*n,i}  + \omega_{*T,i} $ is the ion diamagnetic frequency ($\omega_{*n,i} = -i\frac{cT_{i0}}{Z_iB_0}\mathbf{b_0}\times\frac{\nabla n_{i0}}{n_{i0}}\cdot\nabla$ and $\omega_{*T,i} = -i\frac{c}{Z_iB_0}\mathbf{b_0}\times\nabla T_{i0}\cdot\nabla$), $\eta_{||} = 0.51\frac{m_e\nu_{ei}}{n_{e0}e^2}$ and $\nu_{ei}$ is the electron-ion collision frequency.
The electron perturbed pressure $\delta P_e$ in Eqs. \eqref{vorticity} and \eqref{sound} can be expressed as
\begin{flalign}\label{Pe}
\begin{split}
\delta P_e = \delta n_e T_{e0} + n_{e0}\delta T_e
\end{split},
\end{flalign}
where the electron perturbed temperature $\delta T_e$ is determined by the isothermal condition
\begin{flalign}\label{Te}
\begin{split}
\mathbf{b_0}\cdot\nabla\delta T_e + \frac{1}{B_0}\mathbf{\delta B}\cdot\nabla T_{e0} = 0
\end{split},
\end{flalign}
and the electron perturbed density $\delta n_e$ is calculated through the quasi-neutrality condition
\begin{flalign}\label{poisson}
\begin{split}
e\delta n_e
= Z_i\delta n_i 
\underbrace{+ \frac{c^2}{4\pi V_A^2}\nabla_\perp^2\delta\phi}_{\{Drift\}}
\end{split}.
\end{flalign}
We have omitted the FLR modification on the ion polarization density in Eq. \eqref{poisson}, which is in the higher order of $k_\perp^2\rho_i^2$ compared to other ion FLR terms in Eqs. \eqref{vorticity}, \eqref{ion_pressure} and \eqref{ion_density}.

The electron parallel perturbed velocity $\delta u_{||e}$ in the electron Landau closure term of Eqs. \eqref{ohm_law} and \eqref{sound} are calculated by inverting the Ampere's law
\begin{flalign}\label{Apara}
\begin{split}
en_{e0}\delta u_{||e}
=Z_in_{i0}\delta u_{||i}
\underbrace{ + \frac{c}{4\pi}\nabla_\perp^2\delta A_{||}}_{\{Drift\}}
\end{split},
\end{flalign}
and the thermal ion perturbed temperature $\delta T_i$ in the ion Landau closure term of Eq. \eqref{ion_pressure} can be expressed as
\begin{flalign}\label{Ti}
\begin{split}
\delta T_i = \frac{1}{n_{i0}}\left(\delta P_i - \delta n_iT_{i0}\right)
\end{split}.
\end{flalign}

Eqs. \eqref{vorticity}-\eqref{Ti} form a closed system of Landau-fluid formulation based on the extended drift-ordering, namely, incorporating the ion and electron diamagnetic drifts on an equal footing with $\mathbf{E}\times\mathbf{B}$ drift, as well as keeping ion and electron Landau damping effects and ion FLR. Note that the leading effect of parallel magnetic field perturbation $\delta B_{||}$ has been imposed in the last term on the LHS of Eq. \eqref{vorticity} (i.e., interchange term) through the perpendicular force balance implicitly, which leads to a modification on the magnetic drift (i.e., drift reversal cancellation by replacing $\nabla B_0/B_0$ with $\boldsymbol{\kappa}$) and is shown to be important on KBM \cite{joiner2010, dong2017}, internal kink \cite{brochard2022} and the low frequency Alfv\'en mode (LFAM) identified in DIII-D recently \cite{choi2021}. {\color{black}On the other hand, the perpendicular compressibility associated with $\delta B_{||}$ gives rise to the fast wave branch, which is removed from the Landau-fluid model for avoiding the numerical pollution on the slow mode spectra of physics interest \cite{jardin}.} Moreover, the well-known gyro-viscosity cancellation \cite{hazeltine} has been considered in the derivations of Eqs. \eqref{vorticity}, \eqref{ion_pressure} and \eqref{ion_density} for thermal ions based on the Braginskii model \cite{braginskii1965}. The electron model described by Eqs. \eqref{ohm_law} and \eqref{Te} retains both the adiabatic and convective responses to the electromagnetic fields, which is also consistent with the adiabatic density response in gyrokinetic simulations, namely, Eq. (19) of Ref. \cite{bao2017}.

\section{Numerical method}\label{numerical_method}
MAS solves $\left(\delta\phi, \delta A_{||}, \delta P_i, \delta u_{||i}, \delta n_i\right)$ in general geometry using Landau-fluid model described in section \ref{2}, which applies the finite difference discretization in radial direction and Fourier decompositions in poloidal and toroidal directions. Specifically, MAS first constructs the Boozer coordinates based on the EFIT equilibria, second expands the physical operators in Boozer coordinates and constructs the matrix equation in the form of general linear eigenvalue problem, and third solves the eigenvalues and eigenvectors of the matrix equation. The coordinate system, operator discretization and matrix structure are described in this section.

\subsection{MAS coordinate system: Boozer coordinates}
The general flux coordinates use poloidal magnetic flux $\psi$ as the radial coordinate, and choose proper poloidal angle $\theta$ and toroidal angle $\zeta$ to guarantee the straight field line nature of the grids aligned along $\alpha = q\left(\psi\right)\theta - \zeta$ direction. Boozer coordinates are one special case of general flux coordinates, which additionally have the simple forms of $\mathbf{B_0}$ field in both covariant and contravariant representations besides the straight field line property
\begin{flalign}\label{B_cova}
	\begin{split}
		\mathbf{B_0} = \delta\left(\psi,\theta\right)\nabla\psi + I\left(\psi\right)\nabla\theta + g\left(\psi\right)\nabla\zeta
	\end{split}
\end{flalign}
and
\begin{flalign}\label{B_controva}
	\begin{split}
		\mathbf{B_0} = q\left(\psi\right)\nabla\psi\times\nabla\theta - \nabla\psi\times\nabla\zeta
	\end{split},
\end{flalign}
where $I\left(\psi\right)$, $g\left(\psi\right)$ and $q\left(\psi\right)$ only depend on $\psi$,  and $\delta\left(\psi,\theta\right)$ arises due to the nonorthogonality of $\left(\psi,\theta,\zeta\right)$ coordinates \cite{boozer1981}. Then the Jacobian can be obtained using Eqs. \eqref{B_cova} and \eqref{B_controva} as
\begin{flalign}\label{Jacobian}
	\begin{split}
		J\left(\psi,\theta\right) = \left(\nabla\psi\times\nabla\theta\cdot\nabla\zeta\right)^{-1}
		=\frac{g\left(\psi\right)q\left(\psi\right) + I\left(\psi\right)}{B_0^2}
	\end{split}.
\end{flalign}
The gradient and Laplacian operators in general flux coordinates (including Boozer coordinates) can be expressed as \cite{white}
\begin{flalign}\label{nabla}
	\begin{split}
		\nabla = \frac{\partial }{\partial\psi}\nabla\psi + \frac{\partial }{\partial\theta}\nabla\theta + \frac{\partial }{\partial\zeta}\nabla\zeta
	\end{split}
\end{flalign}
and
\begin{flalign}\label{Laplacian}
	\begin{split}
		\nabla^2 
		=&\frac{1}{J}\frac{\partial}{\partial\psi}\left[
		Jg^{\psi\psi}\frac{\partial}{\partial\psi}
		+Jg^{\psi\theta}\frac{\partial}{\partial\theta}
		+Jg^{\psi\zeta}\frac{\partial}{\partial\zeta}
		\right]
		+\frac{1}{J}\frac{\partial}{\partial\theta}\left[
		Jg^{\psi\theta}\frac{\partial}{\partial\psi}
		+Jg^{\theta\theta}\frac{\partial}{\partial\theta}
		+Jg^{\theta\zeta}\frac{\partial}{\partial\zeta}
		\right]\\
		&+\frac{1}{J}\frac{\partial}{\partial\zeta}\left[
		Jg^{\psi\zeta}\frac{\partial}{\partial\psi}
		+Jg^{\theta\zeta}\frac{\partial}{\partial\theta}
		+Jg^{\zeta\zeta}\frac{\partial}{\partial\zeta}
		\right]
	\end{split},
\end{flalign}
where $g^{\xi_\alpha\xi_\beta} = \nabla\xi_\alpha\cdot\nabla\xi_\beta$ represents the metric tensor element, with $\xi_\alpha$ and $\xi_\beta$ being the $\left(\psi,\theta,\zeta\right)$ dimensions of Boozer coordinates. Considering Eqs. \eqref{B_cova}-\eqref{Laplacian}, the Landau-fluid model in section \ref{2} can be expanded in Boozer coordinates straightforwardly, which fully retains the complex geometry effects in a concise form. 

\begin{figure}[H]
	\center
	\includegraphics[width=0.95\textwidth]{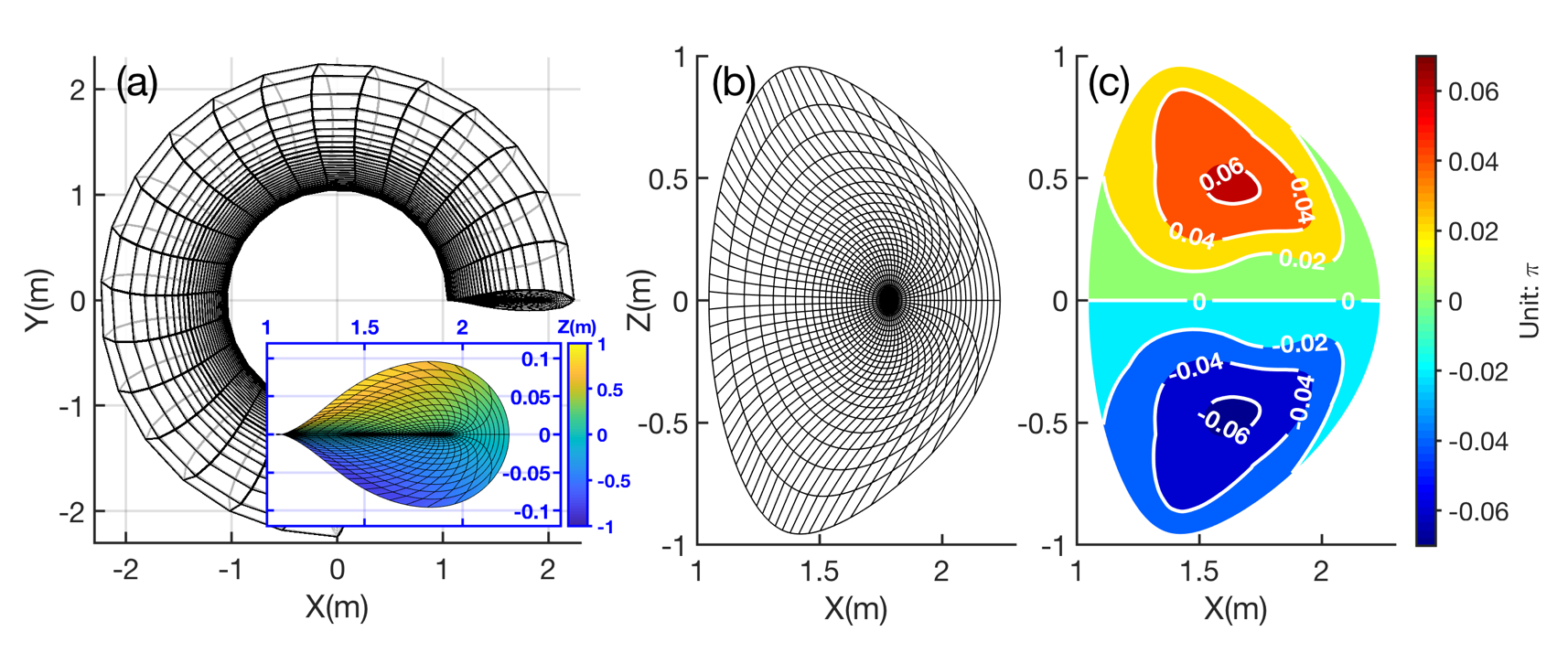}
	\caption{The mesh grids regularly aligned along Boozer coordinates $\psi$, $\theta$, and $\zeta$, which are viewed in the Cartesian coordinate system $\left(X, Y, Z\right)$. (a) The X-Y plane view of mesh grids on the flux surface, and the X-Y plane view of mesh grids in the $\zeta = 0$ poloidal plane where the color denotes $Z$ coordinate. (b) The X-Z plane view of mesh grids in the $\zeta = 0$ poloidal plane. (c) The $\nu\left(\psi, \theta\right)$ difference between Boozer toroidal angle $\zeta$ and cylindrical angle $\phi$.}
	\label{grid_png}	
\end{figure}

We then show the mesh system of MAS in Cartesian coordinate system in figure \ref{grid_png}, which are regularly aligned along $\psi$, $\theta$ and $\zeta$. It is noted that the poloidal plane grids with the same $\zeta$ angle are on a curved surface in figure \ref{grid_png} (a), and the difference between $\zeta$ and the cylindrical angle $\phi$, i.e., $\nu\left(\psi,\theta\right) = \phi-\zeta$, relies on $\psi$ and $\theta$ for axisymmetric tokamak as shown in figure \ref{grid_png} (c). In order to obtain a straight field line system, the choice of $\theta$ coordinate in Eqs. \eqref{B_cova} - \eqref{Jacobian} is also different from the geometric poloidal angle as shown in figure \eqref{grid_png} (b). The contravariant basis vectors $\left(\nabla\psi,\nabla\theta,\nabla\zeta\right)$ satisfy the right-handed rule by Eq. \eqref{Jacobian}, and the convention of basis vector direction in MAS is introduced in Appendix \ref{A5}. In addition, EFIT equilibrium represented in cylindrical coordinates $\left(R,\phi,Z\right)$ is used as input for MAS to construct Boozer coordinates $\left(\psi,\theta,\zeta\right)$, and the mapping algorithm between these two different coordinates is left to a separate work.

\subsection{Operator discretization and matrix structure}
For numerical convenience, we define
\begin{flalign}\label{X_vec}
	\begin{split}
		\mathbf{X} = \left(\widetilde{\delta\phi}, \widetilde{\delta A_{||}}, \widetilde{\delta P_i}, \widetilde{\delta u_{||i}}, \widetilde{\delta n_i}\right)^T =\left(-ic\delta\phi, \delta A_{||}/B_0, \delta P_i, i\delta u_{||i}/B_0, \delta n_i\right)^T
	\end{split}
\end{flalign}
as the unknown vector, then Eqs. \eqref{vorticity}-\eqref{ion_density} can be casted into a matrix equation with considering Eqs. \eqref{Pe}-\eqref{Ti}
\begin{flalign}\label{matrix}
	\begin{split}
		\mathbb{A}\mathbf{X} = \omega\mathbb{B}\mathbf{X}
	\end{split},
\end{flalign}
where $\omega$ is the eigenvalue of frequency, $\mathbb{A}$ and $\mathbb{B}$ are two sparse matrices of operators which are independent of $\omega$. The unknown vector $\mathbf{X}$ is represented in Boozer coordinates as 
\begin{flalign}\label{Xm}
	\mathbf{X}= \sum_m \mathbf{X}_m\left(\psi\right)exp\left(-im\theta +in\zeta\right),
\end{flalign}
where $m$ and $n$ are the poloidal and toroidal mode numbers, i.e., Fourier conjugates of Boozer poloidal angle $\theta$ and toroidal angle $\zeta$. 

\begin{figure}[H]
	\center
	\includegraphics[width=0.9\textwidth]{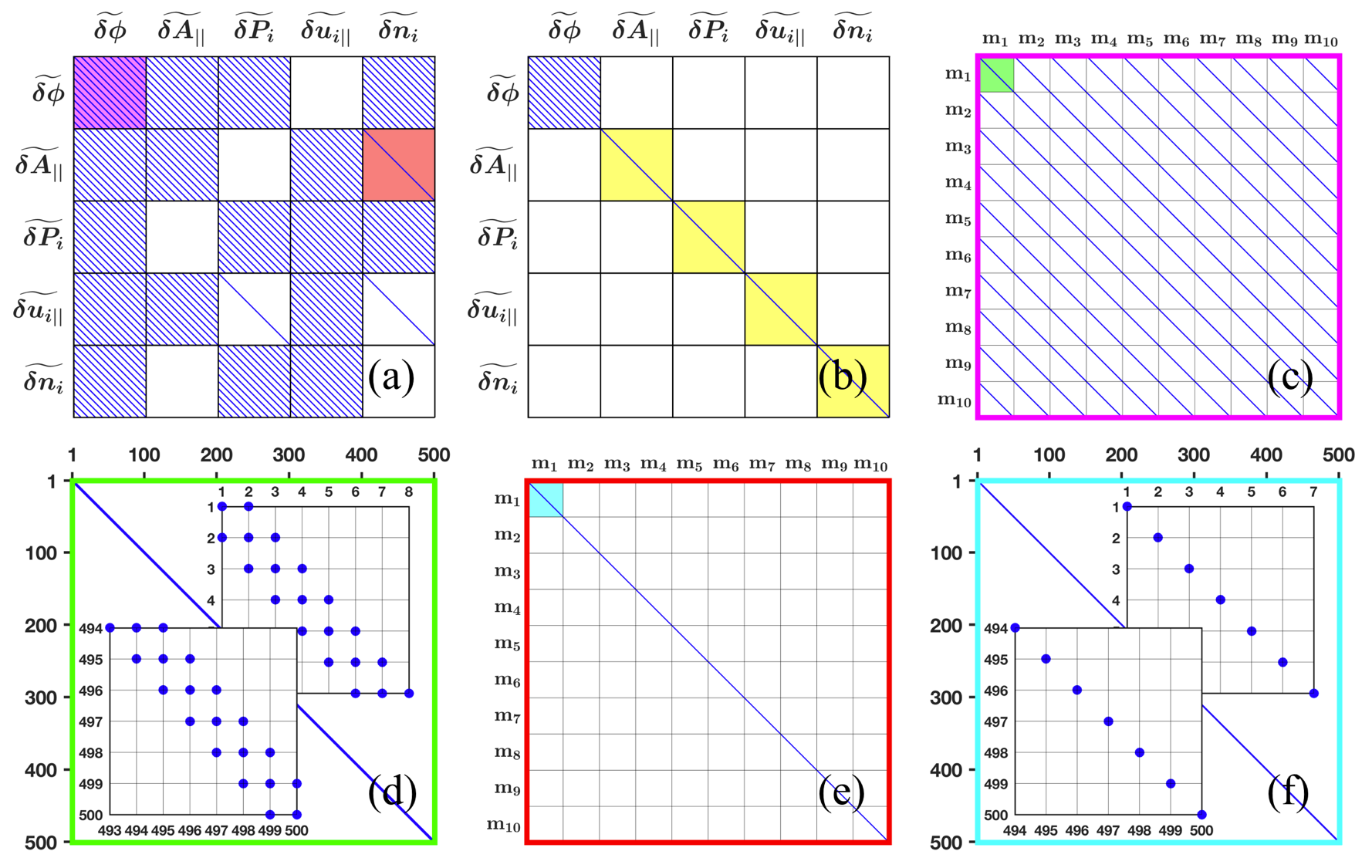}
	\caption{(a) and (b) show the structures of $\mathbb{A}$ and $\mathbb{B}$ matrices respectively, where the blue lines represent the non-zero elements. The $M\times M$ poloidal blocks are shown in panels (c) and (e), which correspond to the purple and red shaded regions in $\mathbb{A}$ respectively. The sparse $R\times R$ dimension submatrices are shown in panels (d) and (f), which correspond to the green and cyan shaded regions in (c) and (e) respectively. The yellow shaded regions in (b) represent the identity submatrices.} 
	\label{mat_png}	
\end{figure}

The structures of matrices $\mathbb{A}$ and $\mathbb{B}$ are shown in figures \ref{mat_png} (a) and (b). Based on the definition of $\mathbf{X}$ in Eqs. \eqref{X_vec} and \eqref{Xm} with $n$ being linearly conserved in an axisymmetric tokamak, it is natural to partition $\mathbb{A}$ and $\mathbb{B}$ into blocks in three levels, i.e., $5\times5$ physics variable blocks in terms of $\left(\widetilde{\delta\phi}, \widetilde{\delta A_{||}}, \widetilde{\delta P_i}, \widetilde{\delta u_{||i}}, \widetilde{\delta n_i}\right)$ in the first level, each physics variable block is further partitioned into $M\times M$ poloidal blocks (where $M$ is the number of $m$-harmonics) in terms of poloidal coupling between $m$-harmonics in the second level, and each poloidal block is a $R\times R$ dimension submatrix (where $R$ is the radial grid number) which represents the radial dependence of operators in the third level. 

Specifically, all orders of poloidal coupling, i.e., $m\pm\Delta m$ with $\Delta m\geq 1$, can be handled in MAS with the constraints $m+\Delta m\leq m_{max}$ and $m-\Delta m>m_{min}$ when the physical operator has $\theta$ dependence, while the poloidal coupling is not necessary (i.e., $\Delta m = 0$) when the physical operator does not rely on $\theta$. The structures of physics variable block with and without poloidal coupling are shown in figures \ref{mat_png} (c) and (e), respectively, which correspond to the purple and red shaded regions in figure \ref{mat_png} (a). 
The radial discretization is incorporated in each poloidal block, for example, figures \ref{mat_png} (d) shows the detailed structure of the green shaded region in figure \ref{mat_png} (c), where finite difference method with second order accuracy (i.e., three radial grids are involved) is applied for discretizing the radially differential operators (i.e., first and second orders derivatives with respect to $\psi$), and the Dirichlet boundary condition is applied at the inner- and outer-radial boundaries. Figure \ref{mat_png} (f) shows the diagonal matrix for operator with zeroth derivative at $\psi$, which corresponds to the cyan shaded region in figure \ref{mat_png} (e). 

{\color{black}The numerical framework of MAS is constructed in MATLAB. Besides implementing the $\mathbb{A}$ and $\mathbb{B}$ matrices for discretized physical operators, we solve eigenvalue $\omega$ and eigenvector $\mathbf{X}$ of Eq. \eqref{matrix} by using eigs in MATLAB, which is efficient for finding solutions in the specific $\omega$ range of interest. For example, the numerical simulation of reversed shear Alfv\'en eigenmode (RSAE) in figure \ref{kink_mode_structure} is performed on a laptop with Intel Core i9 CPU (2.3 GHz, 8-Core), which takes about 3 min with using $R = 513$ radial grids and $M = 17$ coupled $m$-harmonics, and the detailed cost of CPU time includes 140 s for mapping Boozer coordinates based on EFIT equilibrium, 40 s for preparing the physical operator matrices, and 10 s for solving the matrix equation, namely, Eq. \eqref{matrix}.}

\section{Theoretical benchmarks: reductions of Landau-fluid formulation in various limits}\label{3}
In order to delineate the validity regime of present Landau-fluid model in a clear and concise manner, we show the reductions to several well-known reduced models in the interested limits in this section.

\subsection{Ideal full-MHD model}\label{2.2}
Using the ideal MHD ordering, the terms of Landau-fluid model described in section \ref{2} labeled by \{drift\}, \{Ion FLR\}, \{Ion Landau\}, \{Electron Landau\} and \{Resistivity\} are negligibly small and removed, then Eqs. \eqref{vorticity}-\eqref{ion_density} reduce to
\begin{flalign}\label{vorticity-full}
	\begin{split}
		\frac{\partial }{\partial t}\frac{c}{V_A^2}\nabla_\perp^2\delta\phi
		+\mathbf{B_0}\cdot\nabla\left(\frac{1}{B_0}\nabla_\perp^2\delta A_{||}\right) 
		-\frac{4\pi}{c}\boldsymbol{\delta B}\cdot\nabla\left(\frac{J_{||0}}{B_0}\right)
		- 8\pi\left(\nabla\delta P_i + \nabla\delta P_e\right)\cdot\frac{\mathbf{b_0}\times\boldsymbol{\kappa}}{B_0}
		=0
	\end{split},
\end{flalign}
\begin{flalign}\label{ohm_full}
	\begin{split}
		\frac{\partial \delta A_{||}}{\partial t} = -c\mathbf{b_0}\cdot\nabla\delta \phi 
	\end{split},
\end{flalign}
\begin{flalign}\label{ion_pressure_full}
	\begin{split}
		\frac{\partial \delta P_i}{\partial t} 
		+ \frac{c\mathbf{b_0}\times\nabla\delta\phi}{B_0}\cdot\nabla P_{i0} 
		+2\Gamma_{i\perp} P_{i0}c\nabla\delta\phi\cdot\frac{\mathbf{b_0}\times\boldsymbol{\kappa}}{B_0}
		+\Gamma_{i||} P_{i0}\mathbf{B_0}\cdot\nabla\left(\frac{\delta u_{||i}}{B_0}\right)
		=0
	\end{split},
\end{flalign}
\begin{flalign}\label{sound_full}
	\begin{split}
		m_in_{i0}\frac{\partial \delta u_{||i}}{\partial t} 
		= -\mathbf{b_0}\cdot\nabla\delta P_e
		-\frac{1}{B_0}\boldsymbol{\delta B}\cdot\nabla P_{e0}
		-\mathbf{b_0}\cdot\nabla\delta P_i
		-\frac{1}{B_0}\boldsymbol{\delta B}\cdot\nabla P_{i0}
	\end{split},
\end{flalign}
\begin{flalign}\label{ion_density_full}
	\begin{split}
		\frac{\partial \delta n_i}{\partial t} 
		+ \frac{c\mathbf{b_0}\times\nabla\delta\phi}{B_0}\cdot\nabla n_{i0} 
		+ 2cn_{i0}\nabla\delta\phi \cdot\frac{\mathbf{b_0}\times\boldsymbol{\kappa}}{B_0}
		+ n_{i0}\mathbf{B_0}\cdot\nabla\left(\frac{\delta u_{||i}}{B_0}\right)
		=0
	\end{split}.
\end{flalign}
Note that the charge and current densities of ion and electron species are equal to each other due to the one-fluid MHD constraint, then Eqs. \eqref{poisson} and \eqref{Apara} become
\begin{flalign}\label{poisson_mhd}
	\begin{split}
		e\delta n_e
		= Z_i\delta n_i 
	\end{split}
\end{flalign}
and
\begin{flalign}\label{Apara_mhd}
	\begin{split}
		en_{e0}\delta u_{||e}
		=Z_in_{i0}\delta u_{||i}
	\end{split}.
\end{flalign}
Moreover, Eq. \eqref{Pe} and Eq. \eqref{Te} still hold for computing $\delta P_e$, while Eq. \eqref{Ti} is not used in the ideal full-MHD limit, since $\delta T_i$ related term has been dropped in Eq. \eqref{ion_pressure_full}. 

{\color{black}As illustrated in section \ref{2}, an effective $\delta B_{||}$ model is imposed in Eq. \eqref{vorticity-full} by modifying the interchange term using $\boldsymbol{\kappa}$ instead of $\nabla B_0/B_0$, which is important for the quantitative accuracy of low-$n$ MHD instabilities such as $m=n=1$ kink mode \cite{graves2019}. Thus, Eqs. \eqref{vorticity-full} - \eqref{Apara_mhd} together with Eqs. \eqref{Pe} and \eqref{Te} consists of the ideal full-MHD model in this work, which captures important MHD compressibilities including $E\times B$ drift compression in presence of non-uniform magnetic field and parallel sound wave, as well as the leading $\delta B_{||}$ effect of drift reversal cancellation.} Though the potential equations using $\left(\delta\phi, \delta A_{||}\right)$ are very different from the classic ideal full-MHD equations using electromagnetic fields $\left(\mathbf{E},\mathbf{B}\right)$, these two forms are equivalent for full-MHD physics in the regimes of $\omega\ll \Omega_{ci}$ and $k_{||}\ll k_\perp$, and the detailed derivation is shown in Appendix \ref{A3}. Compared to classic $\left(\mathbf{E},\mathbf{B}\right)$ form, the $\left(\delta\phi, \delta A_{||}\right)$ form well separates key physics terms in vorticity equation and greatly simplifies the numerical implementation by using scalar equation rather than the vector equation. Furthermore, the $\left(\delta\phi, \delta A_{||}\right)$ form removes the high frequency fast wave with $\omega = k_\perp V_A\sim\Omega_{ci}$, which is free of spectral pollution in traditional MHD stability computation using $\left(\mathbf{E},\mathbf{B}\right)$ form \cite{jardin}.

It should be noted that $\delta E_{||}$ role is subtle in the framework of ideal full-MHD, on one hand, $\delta E_{||} = 0$ is a typically ideal MHD constraint in Eq. \eqref{ohm_full}, on the other hand, $ Z_in_{i0}\delta E_{||} = -\mathbf{b_0}\cdot\nabla\delta P_e - \left(1/B_0\right)\boldsymbol{\delta B}\cdot\nabla P_{e0} \neq 0$ has been imposed in Eq. \eqref{sound_full}, which gives rise to the ISW as pointed out by J. Freidberg in Ref. \cite{freidberg}. 

\subsection{Reduced-MHD model using slow sound approximation}\label{2.3}
Considering the fact that ISW is not a major concern in the Alfv\'en frequency regime such as for toroidal Alfv\'en eigenmode (TAE) etc., {\color{black}one can further simplify the ideal full-MHD model described in section \ref{2.2} by applying the slow sound approximation \cite{chu1992}, i.e., retaining the $E\times B$ drift compression due to non-uniform magnetic field in Eqs. \eqref{ion_pressure_full} and \eqref{ion_density_full} and effective $\delta B_{||}$ modification on interchange term of Eq. \eqref{vorticity-full} while removing the parallel sound wave by setting $\delta u_{||i} = \delta u_{||e} = 0$.} Eq. \eqref{sound_full} is then removed from the system, and Eqs. \eqref{ion_pressure_full} and \eqref{ion_density_full} become
\begin{flalign}\label{ion_pressure_ssw}
	\begin{split}
		\frac{\partial \delta P_i}{\partial t} 
		+ \frac{c\mathbf{b_0}\times\nabla\delta\phi}{B_0}\cdot\nabla P_{i0} 
		+2\Gamma_{i\perp} P_{i0}c\nabla\delta\phi\cdot\frac{\mathbf{b_0}\times\boldsymbol{\kappa}}{B_0}
		=0
	\end{split}
\end{flalign}
and
\begin{flalign}\label{ion_density_ssw}
	\begin{split}
		\frac{\partial \delta n_i}{\partial t} 
		+ \frac{c\mathbf{b_0}\times\nabla\delta\phi}{B_0}\cdot\nabla n_{i0} 
		+ 2cn_{i0}\nabla\delta\phi \cdot\frac{\mathbf{b_0}\times\boldsymbol{\kappa}}{B_0}
		=0
	\end{split}.
\end{flalign}
It has been shown that the diamagnetic drift convection on MHD $E\times B$ flow is important for the perpendicular momentum balance in the high-$n$ regime that applies drift-ordering \cite{schnack2005}, thus the equivalent term is also retained in the vorticity equation of the reduced-MHD model here
\begin{flalign}\label{ideal reduced}
\begin{split}
	\frac{\partial }{\partial t}\frac{c}{V_A^2}\nabla_\perp^2\delta\phi
	\underbrace{+
		i\omega_{*p,i}\frac{c}{V_A^2}\nabla_\perp^2\delta\phi}_{\{Drift\}}
	+\mathbf{B_0}\cdot\nabla\left(\frac{1}{B_0}\nabla_\perp^2\delta A_{||}\right) 
	-\frac{4\pi}{c}\boldsymbol{\delta B}\cdot\nabla\left(\frac{J_{||0}}{B_0}\right)
	- 8\pi\nabla\left(\delta P_i + \nabla\delta P_e\right)\cdot\frac{\mathbf{b_0}\times\boldsymbol{\kappa}}{B_0}
	=0
\end{split}.
\end{flalign}
Eqs. \eqref{Pe}, \eqref{Te}, \eqref{ohm_full}, \eqref{poisson_mhd} and \eqref{ion_pressure_ssw}-\eqref{ideal reduced} constitute a closed system, which is referred to as the drift reduced-MHD model or the ideal reduced-MHD model depending on whether the second term related to $\omega_{*p,i}$ in Eq. \eqref{ideal reduced} is included or not. One may argue the consistency issue that the diamagnetic drift term is only kept in Eq. \eqref{ideal reduced} while not in Eqs. \eqref{ion_pressure_ssw} and \eqref{ion_density_ssw} in the drift reduced-MHD model, the reason is that diamagnetic drift corrections on $\delta P_e$ and $\delta P_i$ are higher order effects for MHD modes near Alfv\'en frequency, due to the partial cancellation of diamagnetic drifts between ion and electron species in total pressure equation (i.e., for $\delta P_e + \delta P_i$) with considering $T_{e0}\sim T_{i0}$ \cite{hazeltine}. The $\omega_{*p,i}$ term in Eq. \eqref{ideal reduced} can be considered as the lowest order FLR effect \cite{deng2012}, which can affect the AE mode structure and frequency found in recent gyrokinetic simulations \cite{wang2013}. The reduced-MHD model using slow sound approximation is efficient for computing high frequency AEs such as {\color{black}toroidal Alfv\'en eigenmode (TAE)}, as well as validating advanced models with more physics effects.

\subsection{Electrostatic ion-fluid model}\label{2.4}
In the electrostatic limit that $\delta A_{||}\to 0$, the Landau-fluid model in section \ref{2} reduces to the three-field electrostatic ion-fluid model with adiabatic electrons \cite{garbet2001}. We shall further demonstrate the valid regime of the electrostatic approximation, which is useful to differentiate the electrostatic and electromagnetic drift wave instabilities, such as ion temperature gradient mode (ITG) and kinetic ballooning mode (KBM). 

Ignoring the second order FLR term and kink drive in Eq. \eqref{vorticity} and defining $\mathbf{v_d} = \frac{cT_{i0}}{Z_i}\frac{\mathbf{b_0}\times\boldsymbol{\kappa}}{B_0}$ and $\mathbf{v_d}\cdot\nabla = i\omega_{d}$, we have
\begin{flalign}\label{es_vor}
	\begin{split}
		-k_{||}k_\perp^2\delta A_{||} 
		= -\left(\omega-\omega_{*p,i}\right)\frac{c}{V_A^2}k_\perp^2\delta \phi
		+ \frac{8\pi Z_i}{cT_{i0}}\omega_{di}\delta P_i
		- \frac{8\pi e}{cT_{e0}}\omega_{de}\delta P_e
	\end{split},
\end{flalign}
and from Eqs. \eqref{ohm_law}, \eqref{Pe} and \eqref{Te}, $\delta P_e$ can be expressed as 
\begin{flalign}\label{pe_appendix}
	\begin{split}
		\delta P_e = -\left(1-\frac{\omega_{*p,e}}{\omega}\right)\frac{en_{e0}}{T_{e0}}\frac{\omega}{ck_{||}}\delta A_{||} + \frac{en_{e0}}{T_{e0}}\delta \phi
	\end{split},
\end{flalign}
where the electron Landau damping and resistivity effects are ignored, $\omega_{*p,e} = \omega_{*n,e} + \omega_{*T,e}$, $\omega_{*n,e} = -i\frac{cT_{e0}}{q_eB_0}\mathbf{b_0}\times\frac{\nabla n_{e0}}{n_{e0}}\cdot\nabla$ and $\omega_{*T,e} = -i\frac{c}{q_eB_0}\mathbf{b_0}\times\nabla T_{e0}\cdot\nabla$. Then one can express $\delta A_{||}$  in terms of $\delta \phi$ and $\delta P_i$ as
\begin{flalign}\label{ui_apara}
	\begin{split}
		\delta A_{||}
		=ck_{||}\frac{\left(\omega-\omega_{*p,i}\right)\frac{k_\perp^2}{V_A^2} - \frac{8\pi Z_i^2n_{i0}}{c^2T_{i0}}\omega_{di}}{k_{||}^2k_\perp^2 + \frac{8\pi e^2n_{e0}}{c^2T_{e0}}\omega\omega_{de}\left(1-\frac{\omega_{*p,e}}{\omega}\right)}\delta\phi
		-ck_{||}\frac{ \frac{8\pi Z_i^2n_{i0}}{c^2T_{i0}}\omega_{di}}{k_{||}^2k_\perp^2 + \frac{8\pi e^2n_{e0}}{c^2T_{e0}}\omega\omega_{de}\left(1-\frac{\omega_{*p,e}}{\omega}\right)}\frac{\delta P_i}{Z_in_{i0}}
	\end{split}.
\end{flalign}
Meanwhile, considering Eqs. \eqref{ohm_law} and \eqref{Te}, we can write Eq. \eqref{sound} as
\begin{flalign}\label{uipara}
	\begin{split}
		\delta u_{||i} = \underbrace{\frac{Z_i}{m_i}\frac{k_{||}}{\omega}\delta\phi}_{\{I\}} 
		\underbrace{+ \frac{1}{m_in_{i0}}\frac{k_{||}}{\omega}\delta P_i}_{\{II\}}
		 \underbrace{-\frac{Z_i}{cm_i}\left(1-\frac{\omega_{*p,i}}{\omega}\right)\delta A_{||}}_{\{III\}}
	\end{split}.
\end{flalign}
Substituting Eq. \eqref{ui_apara} into Eq. \eqref{uipara}, it is straightforward to compare the electrostatic contributions (denoted by terms \{I\} and \{II\}) with electromagnetic contributions (denoted by term \{III\}) to find out the validity regime of electrostatic approximation. In the limit of $\delta A_{||}\to 0$, we have following relations
\begin{flalign}\label{}
	\begin{split}
		\left(\omega-\omega_{*p,i}\right)^2\ll k_{||}^2V_{A}^2
	\end{split}
\end{flalign}
and 
\begin{flalign}\label{}
	\begin{split}
		\beta_i\ll k_\perp^2\rho_i^2\frac{k_{||}^2v_{thi}^2}{\omega_{di}\left(\omega-\omega_{*p,i}\right)}
	\end{split},
\end{flalign}
namely, when the low frequency condition (much lower than Alfv\'en frequency) and the low $\beta_i$ condition (much smaller than the threshold $\beta_{h} = k_\perp^2\rho_i^2\frac{k_{||}^2v_{thi}^2}{\omega_{di}\left(\omega-\omega_{*p,i}\right)}$) are satisfied simultaneously, $\delta A_{||}$ related terms can be safely ignored without sacrificing much physics accuracy compared to the original electromagnetic Landau-fluid model.

\subsection{Comparison with drift-kinetic model in uniform plasmas}\label{2.5}
In order to evaluate the accuracy of Landau resonance in MAS physics model, it is necessary to compare the analytic dispersion relation in uniform plasmas between Landau-fluid model with drift-kinetic model. Dropping the terms associated to non-uniform plasma equilibrium and ion-FLR effects, Eqs. \eqref{vorticity}-\eqref{Pe} reduce to
\begin{flalign}\label{vorticity-kaw}
	\begin{split}
		&\frac{\partial }{\partial t}\frac{c}{V_A^2}\nabla_\perp^2\delta\phi
		+\mathbf{b_0}\cdot\nabla\left(\nabla_\perp^2\delta A_{||}\right) 
		=0
	\end{split},
\end{flalign}

\begin{flalign}\label{}
	\begin{split}
		\frac{\partial \delta A_{||}}{\partial t} =
		& -c\mathbf{b_0}\cdot\nabla\delta \phi 
		+ \frac{cT_{e0}}{en_{e0}}\mathbf{b_0}\cdot\nabla\delta n_e
		+\frac{cm_e}{e}\sqrt{\frac{\pi}{2}}v_{the}|k_{||}|\delta u_{||e}
	\end{split},
\end{flalign}

\begin{flalign}\label{}
	\begin{split}
		\frac{\partial \delta P_i}{\partial t} 
		+\Gamma_{i||} P_{i0}\mathbf{b_0}\cdot\nabla\delta u_{||i}
		+n_{i0}\frac{2}{\sqrt{\pi}}\sqrt{2}v_{thi}|k_{||}|\delta T_i
		=0
	\end{split},
\end{flalign}

\begin{flalign}\label{}
	\begin{split}
		m_in_{i0}\frac{\partial \delta u_{||i}}{\partial t} 
		= -\mathbf{b_0}\cdot\nabla\delta P_e
		-\mathbf{b_0}\cdot\nabla\delta P_i
		-Z_in_{i0}\frac{m_e}{e}\sqrt{\frac{\pi}{2}}v_{the}|k_{||}|\delta u_{e||}
	\end{split},
\end{flalign}

\begin{flalign}\label{ni-kaw}
	\begin{split}
		\frac{\partial \delta n_i}{\partial t} 
		+ n_{i0}\mathbf{b_0}\cdot\nabla\delta u_{||i} = 0
	\end{split},
\end{flalign}
and
\begin{flalign}\label{Pe-kaw}
	\begin{split}
		\delta P_e = \delta n_e T_{e0}
	\end{split}.
\end{flalign}

Taking the ansatz $e^{-i\omega t+i\mathbf{k}\cdot\mathbf{x}}$ and applying the Fourier transform: $\partial_t = -i\omega t$, $\nabla_\perp = i\mathbf{k_\perp}$ and $\mathbf{b_0}\cdot\nabla = ik_{||}$, the linear dispersion relation of {\color{black}kinetic Alfv\'en wave (KAW)} based on Eqs. \eqref{poisson}-\eqref{Ti}, \eqref{vorticity-kaw}-\eqref{Pe-kaw} is
\begin{flalign}\label{DR-gam}
	\begin{split}
		\left[\frac{\omega^2}{k_{||}^2V_A^2}-1\right]\left[R_e^{LF}\left(\xi_e\right)+\frac{T_{e0}}{T_{i0}}\frac{Z_i^2n_{i0}}{e^2n_{e0}}R_i^{LF}\left(\xi_i\right)\right] 
		= \frac{Z_i^2n_{i0}}{e^2n_{e0}}k_\perp^2\rho_s^2
	\end{split},
\end{flalign}
where $\rho_s = C_s/\Omega_{ci}$, $C_s = \sqrt{T_{e0}/m_i}$, $\Omega_{ci} = \frac{Z_iB_0}{cm_i}$, and $V_A = B_0/\sqrt{4\pi n_{i0}m_i}$. $R_e$ and $R_i$ are the plasma response functions for electron and ion species in Landau-fluid
\begin{flalign}\label{Re}
	\begin{split}
		R_e^{LF}\left(\xi_e\right) = \frac{1}{1-i\sqrt{\frac{\pi}{2}}|\xi_e|}
	\end{split}
\end{flalign}
and
\begin{flalign}\label{Ri}
	\begin{split}
		R_i^{LF}\left(\xi_i\right) = \frac{|\xi_i|+i\frac{2}{\sqrt{\pi}}}{-2\xi_i^2|\xi_i| - i\frac{4}{\sqrt{\pi}}\xi_i^2 + \Gamma_{i||}|\xi_i| + i\frac{2}{\sqrt{\pi}}}
	\end{split},
\end{flalign}
where Eq. \eqref{Re} is only valid in the regime of $\omega< k_{||}v_{the}$ due to the fact that electron inertia term is ignored in the Landau-fluid model.

\begin{figure}[H]
	\center
	\includegraphics[width=0.6\textwidth]{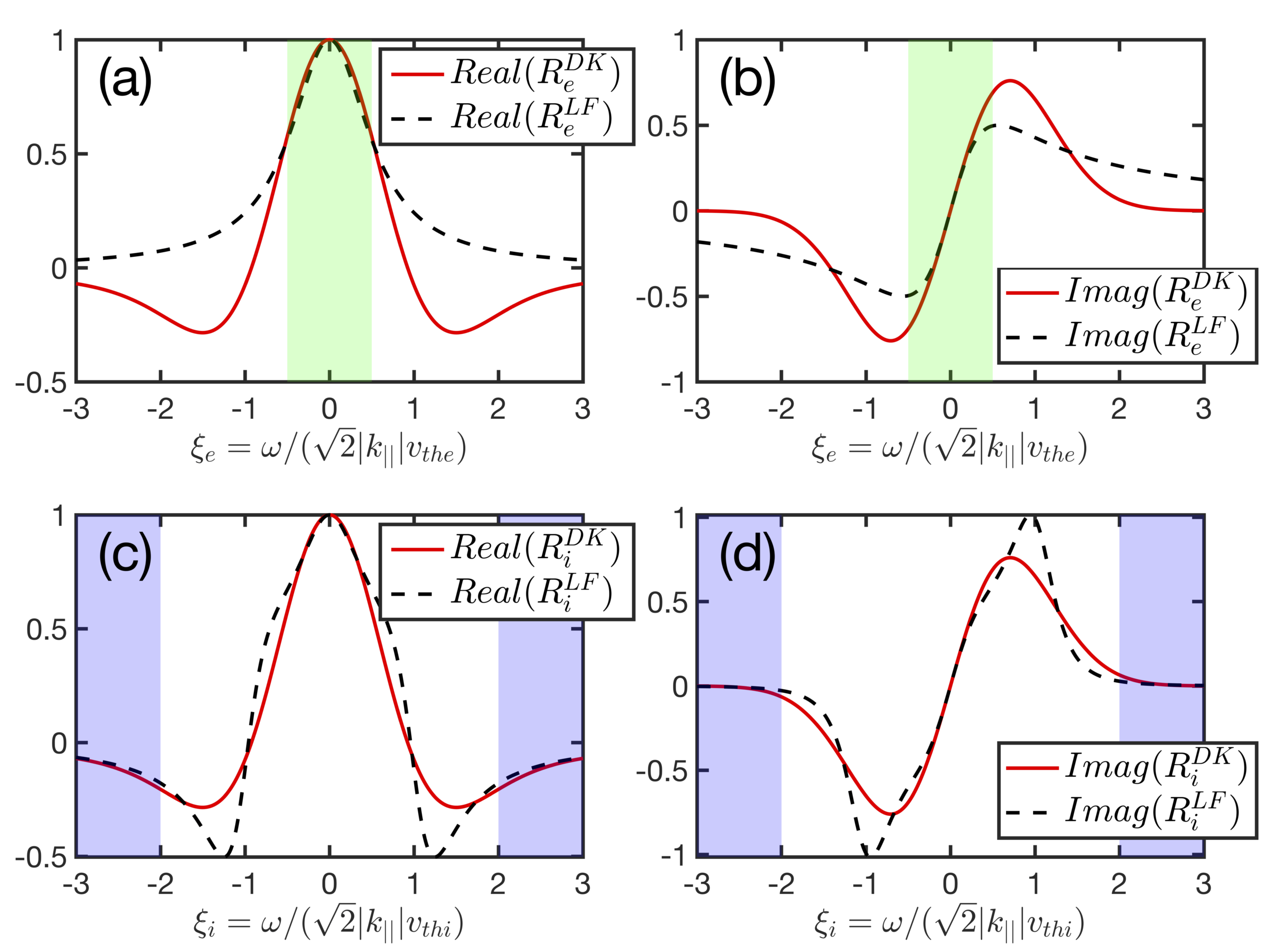}
	\caption{The comparisons of $R_e$ and $R_i$ between Landau-fluid model and drift-kinetic model. The dependences of real and imaginary parts of $R_e$ on $\xi_e$ are shown in panels (a) and (b), and the real and imaginary parts of $R_i$ on $\xi_i$ are shown in panels (c) and (d). The green shaded regions in (a) and (b) indicate the adiabatic electron regime, and the purple shaded regions in (c) and (d) indicate the ion inertia regime, where the two models agree with each other and cover a broad frequency range of typical ion scale waves and instabilities.}
	\label{Rep}	
\end{figure}

For comparison, the linear dispersion relation of drift-kinetic model with both ion and electron species is
\begin{flalign}\label{DR-dk}
	\begin{split}
		\left[\frac{\omega^2}{k_{||}^2V_A^2}-1\right]\left[R_e^{DK}\left(\xi_e\right)+\frac{T_{e0}}{T_{i0}}\frac{Z_i^2n_{i0}}{e^2n_{e0}}R_i^{DK}\left(\xi_i\right)\right] 
		= \frac{Z_i^2n_{i0}}{e^2n_{e0}}k_\perp^2\rho_s^2
	\end{split},
\end{flalign}
where
\begin{flalign}\label{Re-DK}
	\begin{split}
		R_e^{DK}\left(\xi_e\right) = 1 + \xi_e Z\left(\xi_e\right)
	\end{split}
\end{flalign}
and
\begin{flalign}\label{Ri-DK}
	\begin{split}
		R_i^{DK}\left(\xi_i\right) = 1 + \xi_i Z\left(\xi_i\right)
	\end{split},
\end{flalign}
and the derivation of Eq. \eqref{DR-dk} is shown in Appendix \ref{A4}. It is seen that in uniform plasmas, the Landau-fluid dispersion relation Eq. \eqref{DR-gam} has the same form with the drift-kinetic model result Eq. \eqref{DR-dk}, of which differences are only from the definitions of plasma response functions $R_e$ and $R_i$. In order to guarantee the accuracy of Landau resonance, the coefficients in Eqs. \eqref{Re} and \eqref{Ri} have been gauged with benchmarking on analytic dispersion relation in uniform plasmas, for example, $\Gamma_{i||}=3$ should be applied in Eq. \eqref{Ri} in order to match the drift-kinetic result Eq. \eqref{Ri-DK} asymptotically in both $\xi_i\to+\infty$ and $\xi_i\to 0$ regimes \cite{hammett1990}. The comparison of $R_e$ and $R_i$ between Landau-fluid model and drift-kinetic model are shown in figure \ref{Rep}, and one can see the differences are small in the interested ranges, i.e., electron adiabatic regime and ion inertia regime.

\begin{table}[H]
	\caption {Comparison of the involved physics effects between comprehensive Landau-fluid model and simplified models from reductions in various limits.} \label{model_tab} 
	\begin{center}
		\begin{tabular}{ l  l  l  l  l  l  l  l  l }
			\hline
			\rule{0pt}{15pt}
			Model &  \makecell[l]{$E\times B$ drift\\ Compression}& \makecell[l]{Acoustic\\ Compression} & \makecell[l]{Diamagnetic\\ drift} & Ion-FLR& $\delta E_{||}$ &\makecell[l]{Landau\\ Resonance} & $\delta A_{||}$ \\ 
			\hline
			\rule{0pt}{15pt}
			(I) Landau-fluid & Yes & Yes & Ion, electron & Yes & Finite & Ion, electron & Yes \\
			\rule{0pt}{15pt}
			(II) Ideal full-MHD& Yes & Yes & No & No & Zero& No & Yes  \\
			\rule{0pt}{15pt}
			(III) Ideal reduced-MHD&  Yes & No & No & No & Zero & No &Yes  \\
			\rule{0pt}{15pt}
			(IV) Drift reduced-MHD& Yes & No & Ion only & No & Zero & No &Yes  \\
			\rule{0pt}{15pt}
			(V) Electrostatic ion-fluid&  Yes & Yes & Ion only & Yes & Finite & Ion only& No \\
			\rule{0pt}{15pt}
			(VI) \makecell{Uniform drift-kinetic} &  No & Yes & No & No &Finite & Ion, electron& Yes \\
			\hline  
		\end{tabular}
	\end{center}
\end{table}

We list the features of Landau-fluid model and several reductions in Table \ref{model_tab}, of which details are described in sections \ref{2} and \ref{3}. It is seen that the comprehensive Landau-fluid model is constrained by ideal full-MHD model in the long wavelength limit, electrostatic ion-fluid model in the low frequency $\left(\omega\ll\omega_A\right)$ and low-$\beta$ limit, and drift-kinetic model with Landau resonance in the uniform plasma limit, which guarantee the reliability of physics model for diverse physics issues with complex polarization features in realistic experiments.

{\color{black}On the other hand, in order to better delineate the validity regime, the restrictions of MAS Landau-fluid model are clarified by comparing with fully gyrokinetic approach such as LIGKA \cite{lauber2007,Lauber2013}. In LIGKA code, all plasma species including thermal ions, thermal electrons and EP ions are described by using gyrokinetic model, and the important wave-particle resonance and FOW effects are obtained numerically through realistic particle orbit integrals in phase space, which provide accurate kinetic responses in both Alfv\'enic and acoustic frequency ranges. In addition, LIGKA can also solve the SAW and ISW spectra based on the reduced gyrokinetic model that uses well-circulating approximation \cite{bierwage2017}, which is suitable in Alfv\'enic frequency range but suffers error in acoustic frequency range. Compared to fully gyrokinetic approach, the main physics restrictions of MAS model include (i) the EP ions are not considered yet; (ii) the thermal ions are described by three-moment Eqs. \eqref{ion_pressure}, \eqref{sound} and \eqref{ion_density} where the trapped ion and FOW effects are absent and the plasma response deviates from gyrokinetic response in the low frequency regime (i.e., thermal ion magnetic drift frequency $\omega_{di} = -i\frac{cT_{i0}}{Z_i}\frac{\mathbf{b_0}\times\boldsymbol{\kappa}}{B_0}\cdot\nabla $); (iii) isothermal condition is applied for thermal electron model which drops electron temperature gradient drive, and other restrictions are the same with thermal ion.} 

{\color{black}Particularly, with further adding thermal ion diamagnetic drifts on top of the model in section \ref{2.5}, the response function of three-moment closure in slab geometry has been derived by Eq. (13) of Ref. \cite{hammett1990}, which can fit the drift-kinetic response quantitatively, i.e., Eq. (12) of Ref. \cite{hammett1990}, and is consistent with corresponding term in the gyrokinetic dispersion relation that uses well-circulating approximation, e.g., D functions in Eq. (10) of Ref. \cite{zonca1996} and Eq. (1) of Ref. \cite{lauber2009}. Thus, the thermal ion model in MAS can incorporate the diamagnetic drift and parallel ion Landau damping effects being close to gyrokinetic model when the magnetic drift frequency of thermal ion is small such as in the Alfv\'enic frequency range $\omega\sim\omega_A\gg\omega_{di}$, while the thermal ion kinetic particle compression (KPC) terms in gyrokinetic dispersion relation, i.e., N and H functions in Eq. (1) of Ref. \cite{lauber2009}, cannot be fitted by current MAS model, which require to implement more closures associated with $\omega_{di}$ \cite{synder2000} besides the Hammett-Perkins closure deployed already \cite{hammett1990}. In a short summary, MAS Landau-fluid model is more appropriate for computing spectra in the Alfv\'enic frequency range that agrees with gyrokinetic dispersion relation quantitatively, while the error associated with $\omega_{di}$ is amplified in the acoustic frequency range where the parallel ion Landau damping is still qualitatively valid.} 

\section{Numerical benchmarks}\label{4}
In order to present MAS capabilities on addressing certain plasma problems and verify the numerical implementation, we carry out MAS benchmark simulations of typical plasma waves and instabilities in this section.

\begin{figure}[H]
	\center
	\includegraphics[width=0.9\textwidth]{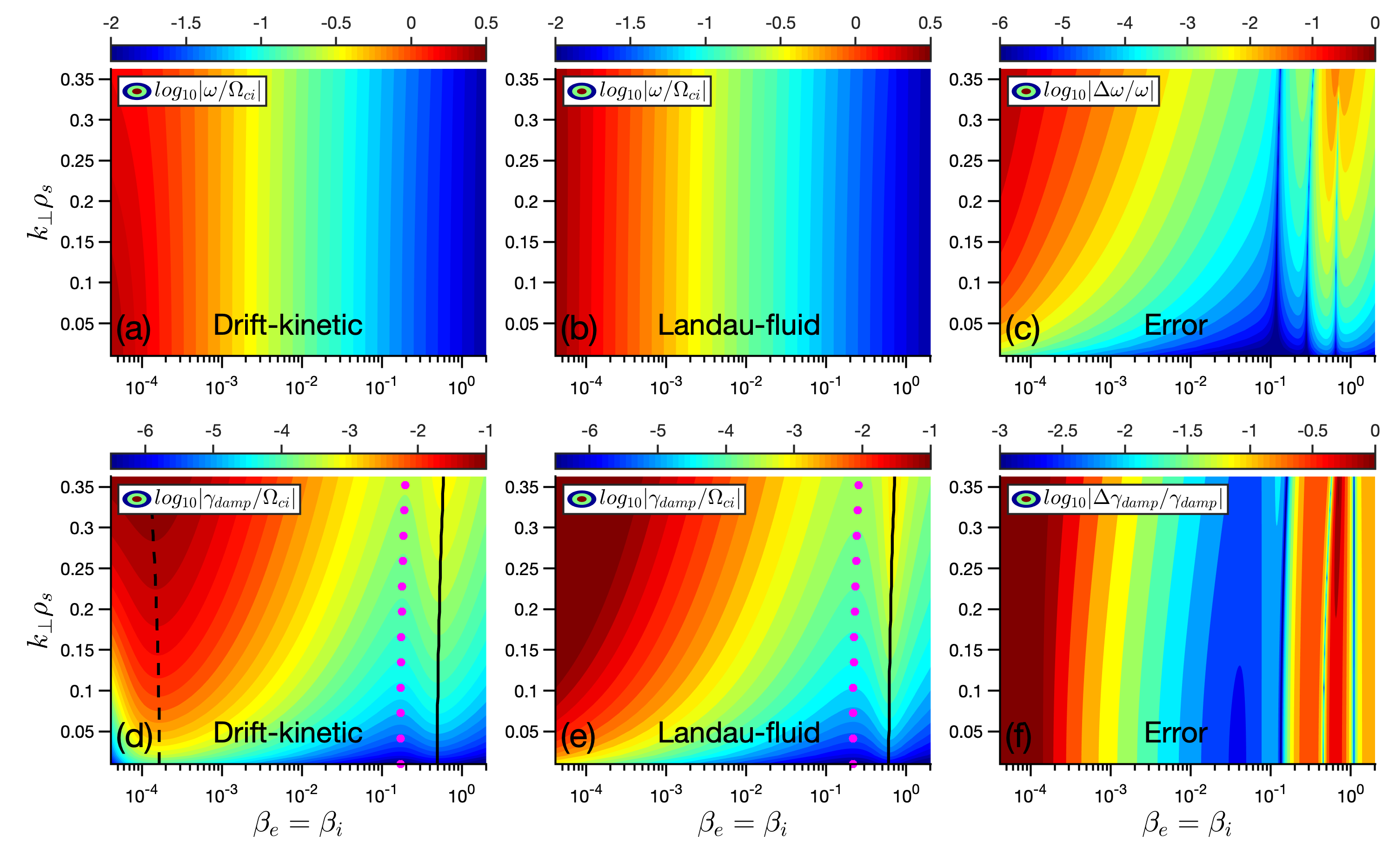}
	\caption{The drift-kinetic solution of KAW (a) frequency and (d) damping rate, and MAS simulations of KAW (b) frequency and (e) damping rate based on the Landau-fluid model. (c) and (f) are the relative errors of frequency and damping rate between drift-kinetic theory and MAS simulations, respectively.}
	\label{kaw_dr_png}	
\end{figure}

\subsection{Normal modes: KAW and ISW with Landau damping}\label{4.1}
There are two branches of normal modes in the linear dispersion relation of drift-kinetic model from Eq. \eqref{DR-dk} : one is the high frequency solution of KAW with $\omega\approx k_{||}V_A\sqrt{1+k_\perp^2\rho_s^2}$, and the other one is low frequency solution of ISW with $\omega\approx k_{||}C_s$, which is solved from $R_e^{DK}\left(\xi_e\right)+\frac{T_{e0}}{T_{i0}}\frac{Z_i^2n_{i0}}{e^2n_{e0}}R_i^{DK}\left(\xi_i\right) \approx0$ with $\xi_e\ll1$ and $\xi_i\gg 1$ \cite{lin2001}. We compare the KAW and ISW solutions from MAS simulations with drift-kinetic theory, in order to verify that the Landau-fluid model can treat the kinetic ion and electron responses accurately in uniform plasmas. The ISW branch is mostly determined by the response functions in Eqs. \eqref{Re}, \eqref{Ri}, \eqref{Re-DK} and \eqref{Ri-DK}, which have been compared in figure \eqref{Rep} with good agreements in the electron adiabatic regime ($\xi_e\ll1$) and ion inertia regime ($\xi_i\gg 1$), thus it has proved that the Landau-fluid model is valid for ISW, and we don't show the repetitive numerical simulations of ISW here. In the following part, we benchmark MAS simulation results of KAW branch against the drift-kinetic theory.

Two species cylindrical plasma with uniform magnetic field in the axial direction is applied in the simulation, which consists of proton ions and electrons. Following equilibrium and perturbation parameters are fixed in the benchmarks: the magnetic field $B_0 = 1.0T$, ion and electron temperatures $T_{i0} = T_{e0}= 1.0keV$, and the parallel wave vector $k_{||} = n/R_0 = 4.0m^{-1}$ (where $n$ is the axial mode number, and the cylinder axial length is $2\pi R_0$ with $R_0 = 1.0m$). Then we scan the plasma density in the range of $n_{i0}=n_{e0} \in \left[1.0\times 10^{11}cm^{-3},5.0\times 10^{15}cm^{-3}\right]$, and $k_\perp\rho_s$ in the range of $\left[0.01,0.36\right]$. The MAS simulation results of KAW frequency and damping rate are shown in figures \ref{kaw_dr_png} (b) and (e), which agree well with drift-kinetic theory solutions (figures \ref{kaw_dr_png} (a) and (d)) in most $\beta-k_\perp\rho_s$ domain. It is also seen from figure \ref{kaw_dr_png} that the KAW damping rate has a much more sensitive dependence on $k_\perp\rho_s$ compared to the frequency.

The KAW damping rate varies non-monotonically with $\beta$ from both drift-kinetic theory and MAS simulation, which is explained as follows. The KAW parallel phase velocity can be approximated as, i.e., $v_p=\omega/k_{||}\approx V_A\sqrt{1+k_\perp^2\rho_s^2}$, then the ratios of $v_p$ to ion and electron thermal velocities, which determine the Landau resonance strength, can be estimated by
\begin{flalign}\label{xii}
	\begin{split}
	\xi_i = \frac{v_p}{\sqrt{2}v_{thi}} 
	= \sqrt{\frac{1}{\beta_i}}\sqrt{1+k_\perp^2\rho_s^2}
	\end{split}
\end{flalign}
and
\begin{flalign}\label{xie}
	\begin{split}
		\xi_e = \frac{v_p}{\sqrt{2}v_{the}} 
	= \sqrt{\frac{1}{\beta_e}}\sqrt{\frac{m_e}{m_i}}\sqrt{1+k_\perp^2\rho_s^2}
	\end{split}.
\end{flalign}

Eqs. \eqref{xii} and \eqref{xie} indicate that the ion and electron Landau dampings favor high- and low-$\beta$ plasma conditions, respectively (i.e., $\beta_i\sim1$ for $\xi_i\sim 1$, and $\beta_e\sim m_e/m_i$ for $\xi_e\sim 1$). In figure \ref{kaw_dr_png} (d) of drift-kinetic results, the dashed and solid lines correspond to the local maximal damping of KAW dominated by electron and ion species respectively, and the magenta circles mark the weakly damped region (i.e., a valley of damping rate variation along $\beta$) due to the lack of resonance condition $k_{||}v_{thi}\ll\omega\ll k_{||}v_{the}$. When $\beta_e$ gets close to and smaller than $m_e/m_i$, the high frequency solution of Eq. \eqref{DR-dk} is no longer KAW (which assumes electron adiabatic response dominates) and becomes the inertia Alfv\'en wave with parallel phase velocity being greater than $v_{the}$, as indicated by the region on the left hand side of the dashed line in figure \ref{kaw_dr_png} (d). There are relatively large differences on both frequency and damping rate in this low-$\beta$ regime ($\beta_e\sim m_e/m_i$) by comparing MAS simulation results (figures \ref{kaw_dr_png} (b) and (e)) and drift-kinetic theory (figures \ref{kaw_dr_png} (a) and (d)), because we drop the electron inertia term in Eq. \eqref{ohm_law} of Landau-fluid model and thus remove the trivial inertia Alfv\'en wave physics. As we expected, MAS simulations of KAW frequency and damping rate are consistent with drift-kinetic theory in the $\beta$ regime of nowadays tokamak (i.e., $\beta_e \sim \beta_i \in \left[0.001, 0.1\right]$), while the extremely low- and high-$\beta$ regimes with relatively large discrepancies are out of tokamak plasma study scope. 

{\color{black}Regarding to the practical applications, MAS is able to simulate the mode conversion processes of KAWs \cite{hasagawa1976} in a wide $\beta$-domain where both electron and ion Landau damping effects are properly taken into account. Furthermore, although ion Landau damping is much weaker than electron Landau damping for KAWs in tokamak plasmas with $\beta\sim0.01$, the KAWs can form the radially bounded states due to geometry effect, such as kinetic beta-induced Alfv\'en eigenmode (KBAE) due to geodesic magnetic curvature, which suffers much stronger ion Landau damping in the regime of $\omega\sim\omega_{ti}$ ($\omega_{ti} = \sqrt{2T_{i0}/m_i}/q/R_0$) \cite{zonca1996,Lauber2013}.}

\subsection{Low-$n$ MHD instability: ideal internal kink mode}\label{4.2}
In section \ref{2.2}, we have shown that MAS Landau-fluid model can faithfully cover the ideal full-MHD physics except for fast wave. However, the MHD reduction of Landau-fluid model still applies the separate heat ratios for ion and electron species, which differs from the one-fluid MHD treatment that uses single heat ratio in Eqs. \eqref{mhd-ohm} and \eqref{mhd-vorticity}-\eqref{mhd-deltaP} of appendix \ref{A3}. The relation of heat ratios between one-fluid MHD and Landau-fluid reduction in section \ref{2.2} can be deduced from Eqs. \eqref{Pe}, \eqref{Te}, \eqref{ohm_full}-\eqref{Apara_mhd} as
\begin{flalign}\label{heatratio}
	\begin{split}
	    \Gamma P_0 = P_{e0} + \Gamma_{i\perp} P_{i0} = P_{e0} + \Gamma_{i||} P_{i0} 
	\end{split},
\end{flalign}
where $P_0 = P_{e0} + P_{i0}$. In Eq. \eqref{heatratio}, the unity heat ratio for electron is consistent with its isothermal assumption in Landau-fluid model, and the MHD isotropic condition requires $\Gamma_{i\perp} = \Gamma_{i||}$. Thus we can adjust $\Gamma_{i\perp}$ and $\Gamma_{i||}$ synchronously in MAS simulation to represent an effective $\Gamma$ of one-fluid MHD.
\begin{figure}[H]
	\center
	\includegraphics[width=0.4\textwidth]{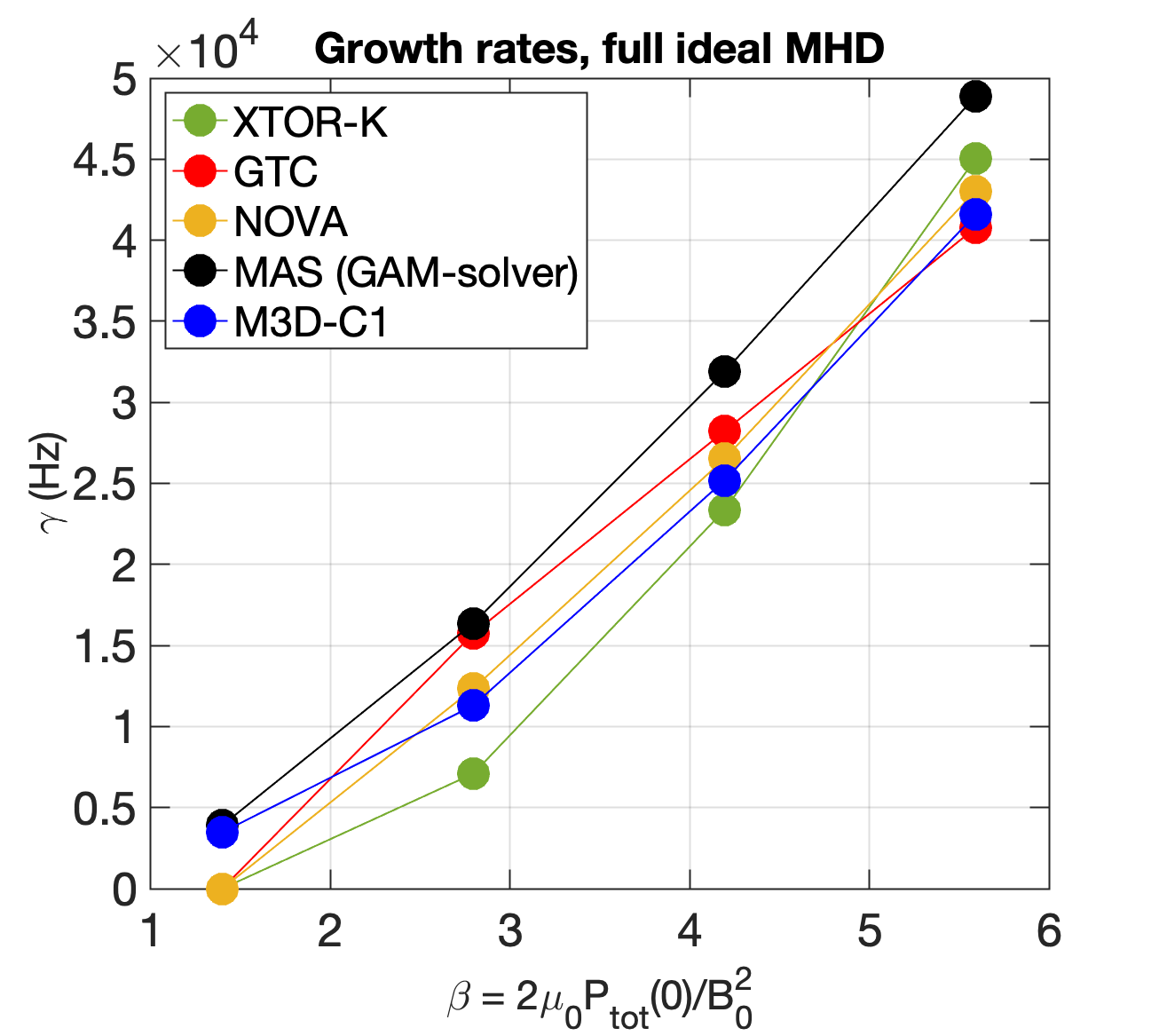}
	\caption{The comparison of internal kink growth rate between different codes in ideal full-MHD limit for DIII-D shot \#141216 at t =1750ms. Good agreement is obtained between codes using kinetic-MHD hybrid and gyrokinetic models. {\color{black}Figure data is adapted from Ref. \cite{brochard2022}. (Note that MAS joins this benchmark study with the former name 'GAM-solver'. ) Copyright 2022 IAEA.}}
	\label{kink_benchmark}	
\end{figure}
\begin{figure}[H]
	\center
	\includegraphics[width=0.5\textwidth]{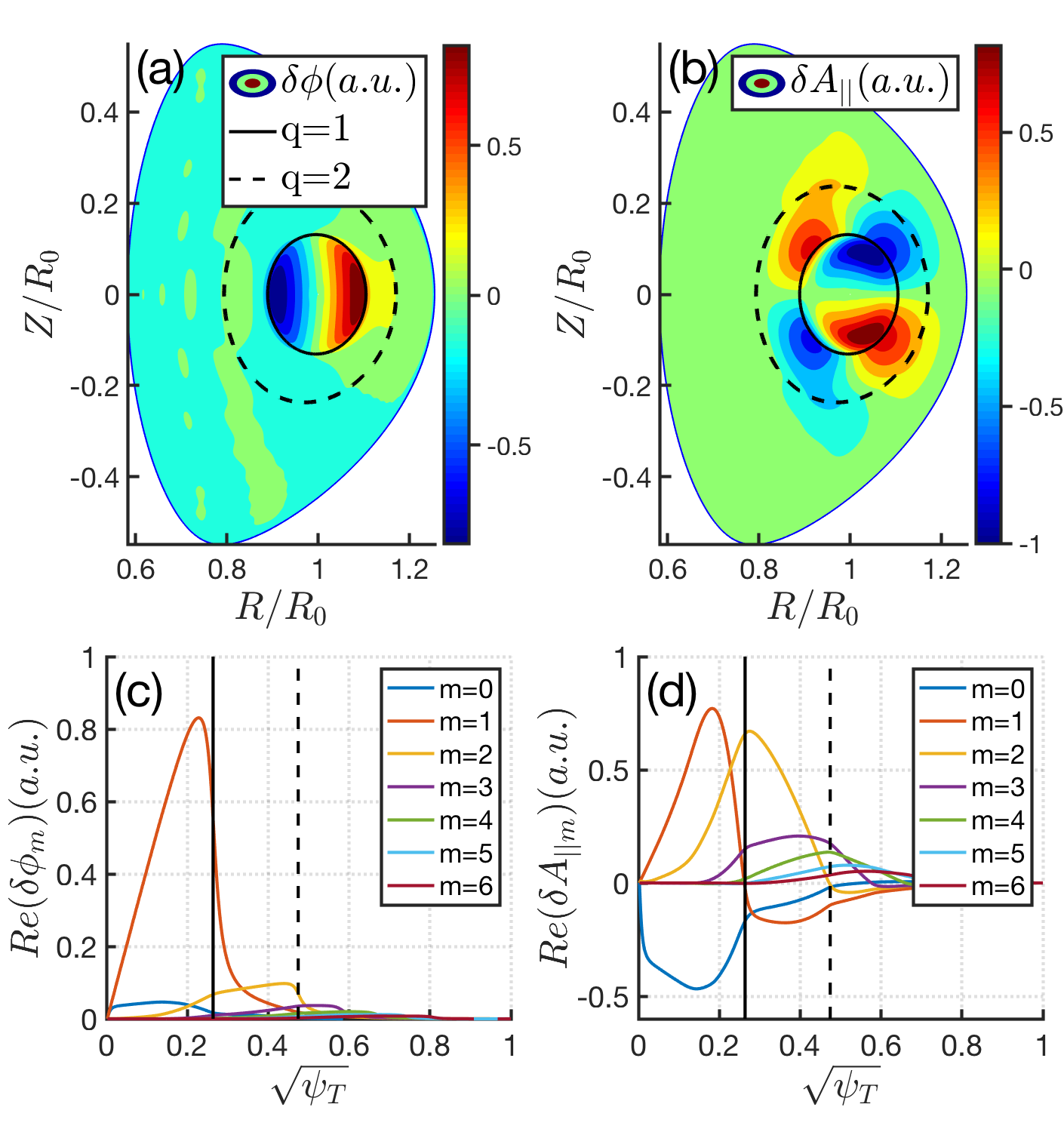}
	\caption{The poloidal mode structures of (a) electrostatic potential $\delta\phi$ and (b) parallel vector potential $\delta A_{||}$. The radial profile of poloidal harmonics of (c) $\delta\phi$ and (d) $\delta A_{||}$. The black solid and dashed lines denote $q = 1$ and $q = 2$ rational surfaces, respectively.}
	\label{kink_mode_structure}	
\end{figure}
MAS has been benchmarked with other kinetic-MHD hybrid and gyrokinetic codes for $m=n=1$ internal kink mode in DIII-D shot \#141216 \cite{brochard2022}, and the comparison of kink growth rate is shown in figure \ref{kink_benchmark}. Besides the excellent agreement of cross-code comparison with the nonideal effects being suppressed in all benchmarking codes, the benchmark shows that kink growth rate scales almost linearly with $\beta$, which is due to the combination of global pressure gradient drive and local current gradient drive at the $q=1$ surface. The mode structures of electrostatic potential $\delta\phi$ and parallel vector potential $\delta A_{||}$ are shown in figure \ref{kink_mode_structure}, $\delta\phi$ is dominant by $m=1$ poloidal harmonic, while $\delta A_{||}$ exhibits the large $m=0$ and $m=2$ sidebands being comparable to the $m=1$ primary poloidal harmonic, which is caused by the ideal MHD constraint (i.e., $ -\mathbf{b_0}\cdot\nabla\delta\phi -\left(1/c\right)\partial_t\delta A_{||} = 0$). We then perform the $\beta$ scan for different $\Gamma$ values as shown in figure \ref{kink_growth_rate}, it is found that the kink growth rates almost remain the same between $\Gamma = 1$ and $\Gamma = 5/3$. {\color{black} However, when we remove all plasma compressibilities by setting $\Gamma = 0$, the kink growth rate increases nearly by a factor of two compared to the finite-$\Gamma$ cases, which indicates the importance of finite-$\beta$ effect on internal kink mode that requires to keep $E\times B$ drift compression due to non-uniform magnetic field, parallel sound wave and $\delta B_{||}$ modification on MHD interchange drive simultaneously, and recent full-MHD theory and simulation also draw similar conclusion \cite{graves2019, pamela2020}.} In a short summary, MAS is capable of simulating macroscopic MHD instability with Dirichlet boundary condition in the plasma edge, and both ideal full-MHD model and comprehensive Landau-fluid model with bulk plasma kinetic effects can be applied.

\begin{figure}[H]
	\center
	\includegraphics[width=0.4\textwidth]{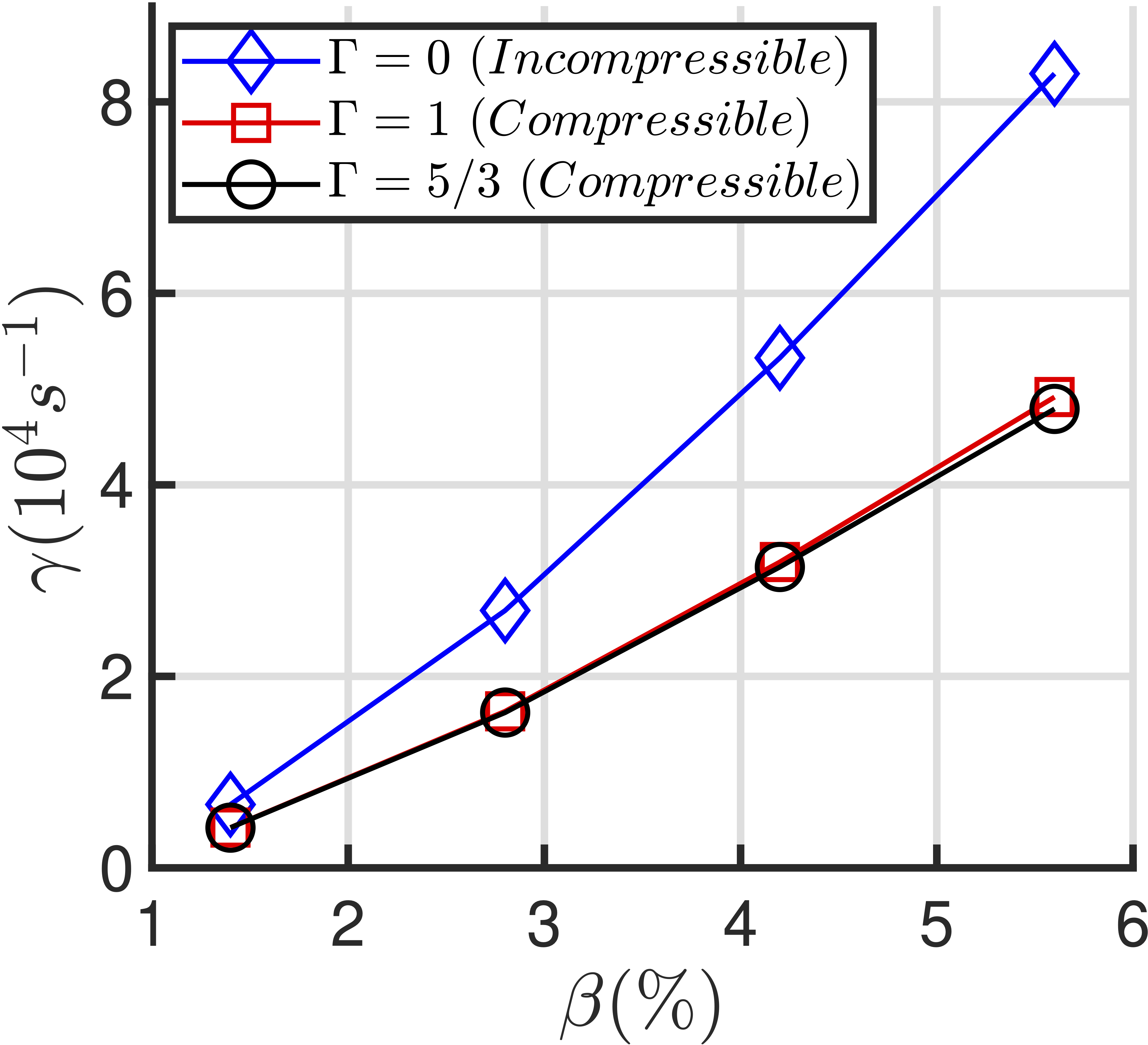}
	\caption{The kink growth rate dependence on the ratio of the plasma pressure to the magnetic pressure. The blue diamond line, red square line and black circle line represent the results using heat ratios of $\Gamma = 0$, $\Gamma = 1$ and $\Gamma = 5/3$, respectively.}
	\label{kink_growth_rate}	
\end{figure}

\subsection{High-$n$ drift wave instabilities: ITG and KBM}\label{4.3}
The MAS simulations of drift wave instabilities are carried out using the Cyclone Base Case (CBC) parameters \cite{dimits2000}. The $s-\alpha$ like concentric circular geometry is applied with $R_0 = 83.5cm$ and $a=0.36R_0$, and the $q = 1.4$ rational surface is located at $r = 0.5a$ with magnetic shear $s=0.78$. The proton thermal ion ($Z_i = e$) temperature profile is equal to the thermal electron. At $r = 0.5a$, $T_{i0} = T_{e0} = 2223eV$, $R_0/L_{Ti} = R_0/L_{Te} = 6.9$, $R_0/L_{ni} = R_0/L_{ne} = 2.2$, and $\eta_i =L_{ni}/L_{Ti}= 3.14$, where $L_T$ and $L_n$ are the scale lengths of temperature and density, respectively. The magnetic field strength on the magnetic axis is $B_a = 2.01T$. The toroidal mode number $n =10$ is chosen in the simulations corresponding to $k_\theta\rho_s = 0.22$ at $r = 0.5$, and the poloidal harmonics $m\in[8,24]$ are kept corresponding to the resonant perturbations on each rational surfaces in the radial domain. To verify the ITG and KBM dispersion relations with former gyrokinetic simulations \cite{dong2017}, we fix the plasma temperature and vary the density for $\beta_e$ scan in the range of $\beta_e \in\left[0, 0.02\right]$. 

It should be pointed out that the thermal ion KPC term has complex dependence on $\left(\omega,\mathbf{k}\right)$ in the first-principle gyrokinetic model \cite {zonca1996,lauber2009,bierwage2017}, and the effective heat ratios $\Gamma_{i\perp,||}\left(\omega,\mathbf{k}\right)$ are no longer constant. Here, we focus on presenting MAS simulation capability of ITG and KBM in Landau-fluid framework, and the constant $\Gamma_{i||}$ and $\Gamma_{i\perp}$ are used for all cases in $\beta_e$ scan, of which dependences on $\left(\omega,\mathbf{k}\right)$ require more detailed derivations and are out of the paper scope. Considering the particle parallel dynamics is much faster than the perpendicular one in the anisotropic magnetized plasmas, $\Gamma_{i||}$ and $\Gamma_{i\perp}$ are generally different from each other, $\Gamma_{i||} = 3$ is chosen for quantitative correctness of ion parallel Landau damping \cite{hammett1990} as confirmed in section \ref{2.5}, and we apply the commonly used heat ratio values in the perpendicular direction, i.e., $\Gamma_{i\perp} = 1$ and $\Gamma_{i\perp} = 5/3$, for following simulations of ITG and KBM.
 \begin{figure}[H]
 	\center
 	\includegraphics[width=0.5\textwidth]{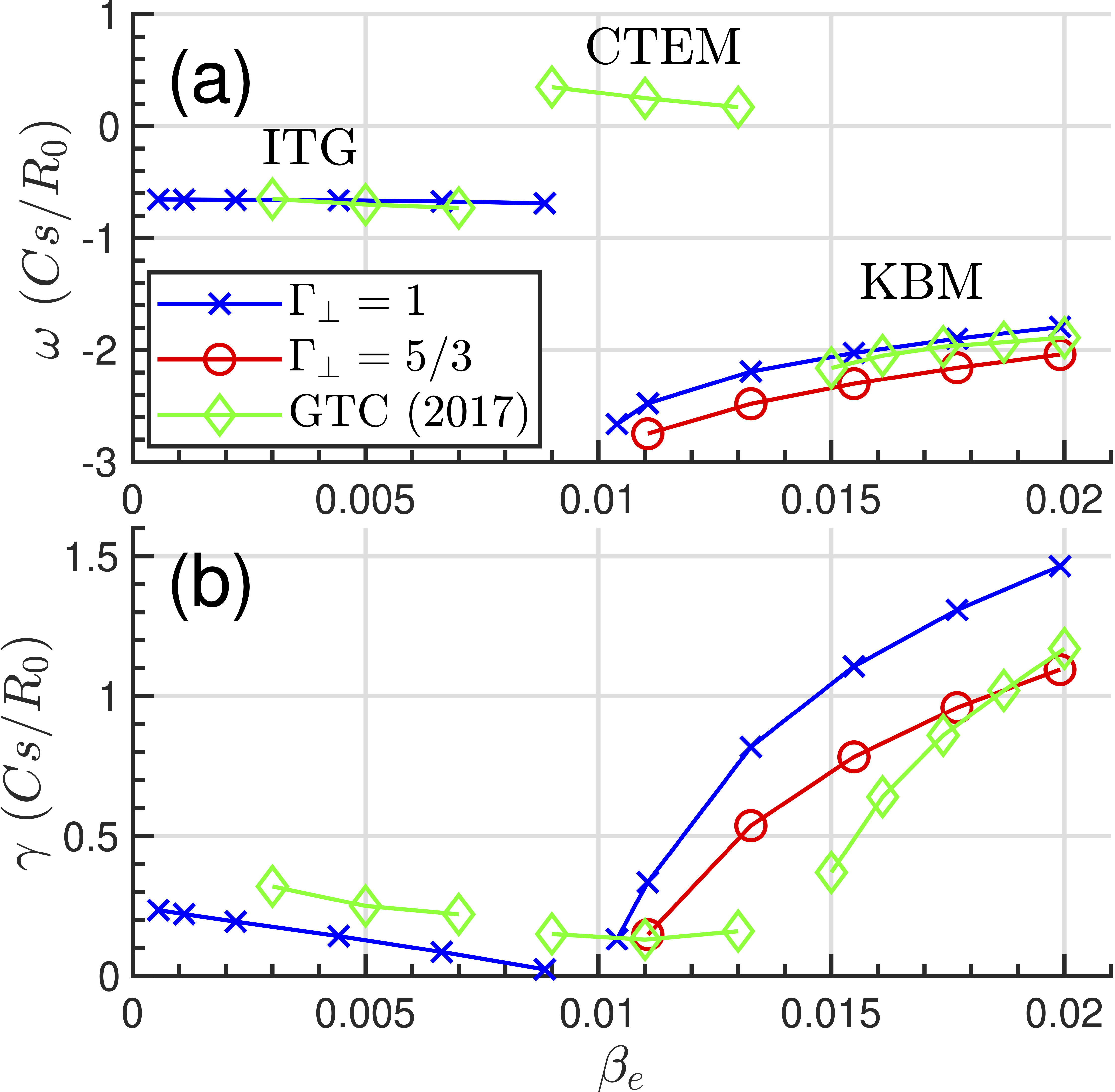}
 	\caption{Comparison of $k_\theta\rho_s = 0.22$ ($n=10$) mode between MAS and GTC simulations using CBC equilibrium with different $\beta_e$ values \cite{dimits2000}. (a) Real frequency and (b) growth rate dependences on $\beta_e$.}
 	\label{dr}	
 \end{figure}
 As shown in figure \ref{dr} (a), MAS simulations using $\Gamma_{i\perp} = 1$ agree well with GTC first-principle results on both ITG and KBM frequencies, while the absence of collisionless trapped electron mode (CTEM) branch in MAS is due to the fact that trapped electron dynamics is not included in Landau-fluid model yet. Meanwhile, the finite-$\beta$ stabilization effect on ITG and the onset of KBM are observed in $\beta_e$ scan with $\Gamma_{i\perp} = 1$ as shown in figure \ref{dr} (b), which are consistent with gyrokinetic simulations \cite{dong2017}. For $\Gamma_{i\perp}=5/3$ cases, only the KBM branch of $\omega_{KBM}\gg\omega_{di}$ is unstable, while the ITG branch of $\omega_{ITG}\sim\omega_{di}$ becomes stable, because $\Gamma_{i\perp}>1$ amplifies $\omega_{di}$ related term in Eq. \eqref{ion_pressure} which effectively stabilizes the low frequency ITG mode.  {\color{black}There are relatively large growth rate differences between MAS and GTC results, which are attributed to trapped electron effects on ITG \cite{lin2007} and KBM \cite{cheng2004}, and the fact that thermal ion KPC term in gyrokinetic dispersion relation \cite{zonca1996,lauber2009} cannot be fitted by present Landau-fluid model in MAS as illustrated in section \ref{3}.} 
 
\begin{figure}[H]
	\center
	\includegraphics[width=1\textwidth]{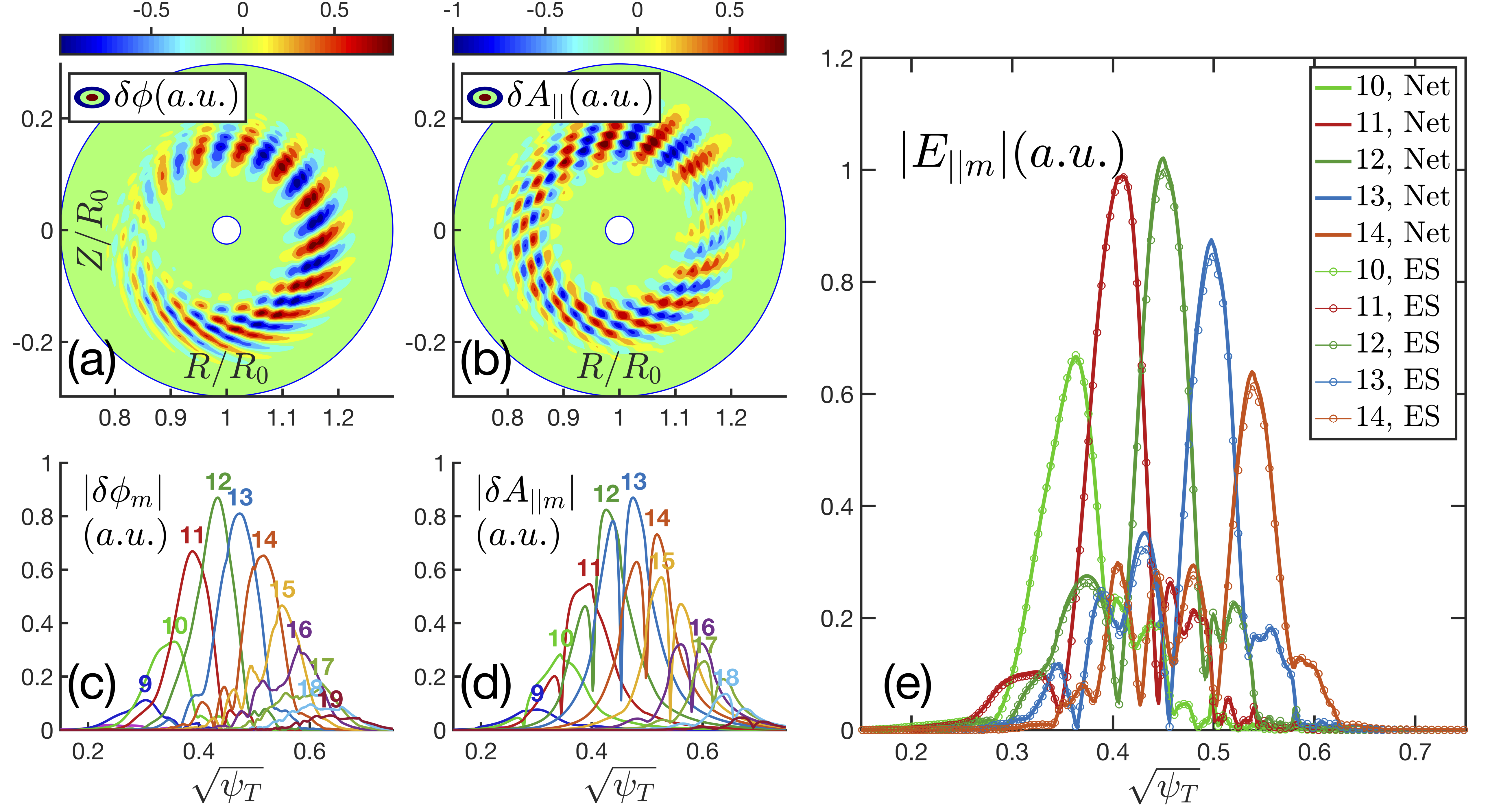}
	\caption{The $k_\theta\rho_s = 0.22$ ($n=10$) ITG mode structures with $\beta_e = 0.0022$ and $\Gamma_{i\perp} = 1$. (a)-(d): The poloidal contour plots and poloidal harmonic radial plots of electrostatic potential $\delta \phi$ and parallel vector potential $\delta A_{||}$. (e) The poloidal harmonic radial plots of parallel electric field $E_{||}$, where the thin lines with circle represent $E_{||}^{ES}=-\mathbf{b_0}\cdot\nabla\delta\phi$ and the thick lines represent $E_{||}^{Net}= -\mathbf{b_0}\cdot\nabla\delta\phi-\left(1/c\right)\partial_t \delta A_{||}$.}
	\label{itg_ms}	
\end{figure}

\begin{figure}[H]
	\center
	\includegraphics[width=1\textwidth]{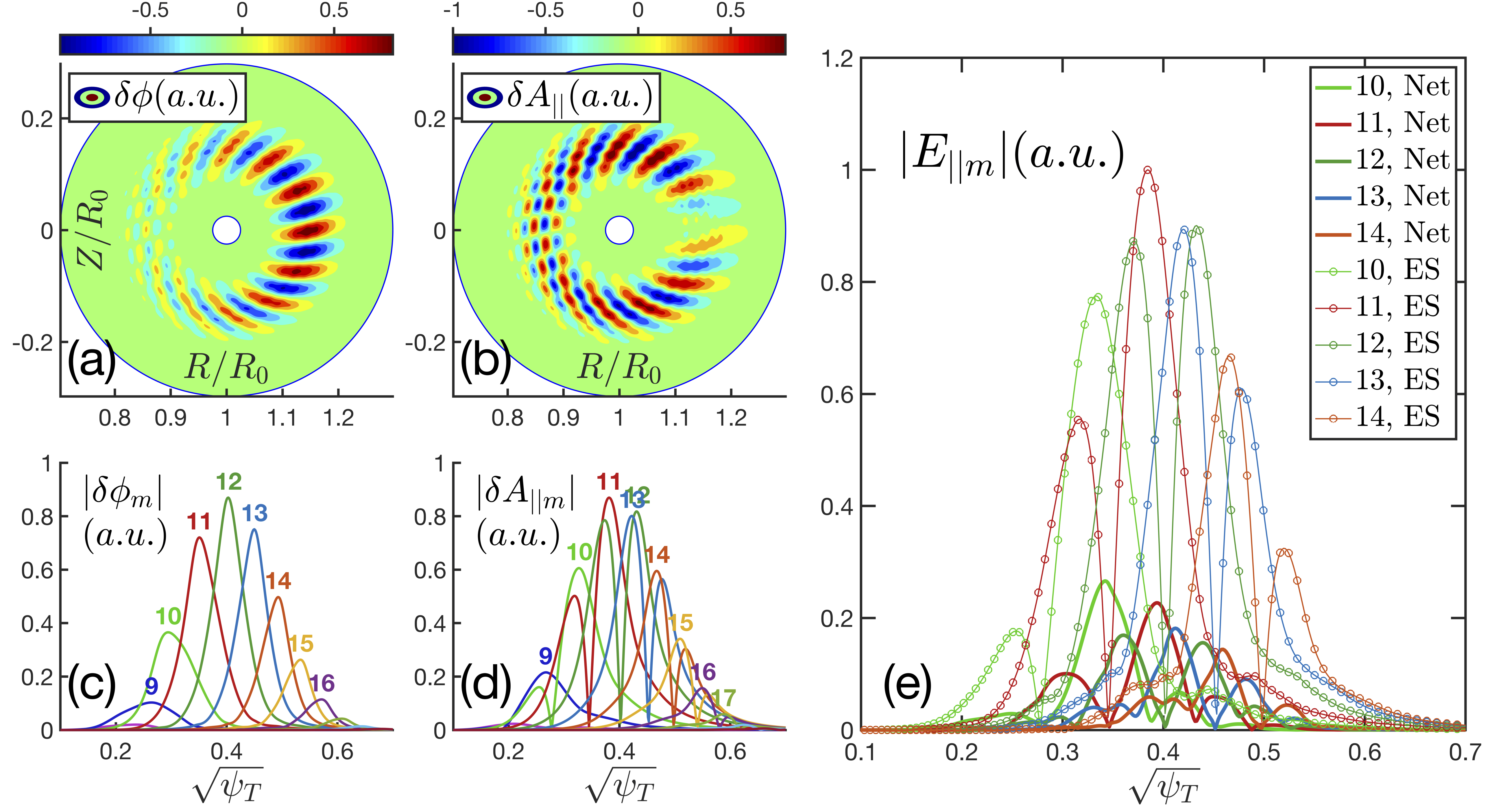}
	\caption{The $k_\theta\rho_s = 0.22$ ($n=10$) KBM mode structures with $\beta_e = 0.02$ and $\Gamma_{i\perp} = 1$. The captions of (a)-(e) are the same with figure \ref{itg_ms}.}
	\label{kbm_ms}	
\end{figure}

 Next we show the mode structures of $\delta\phi$ and $\delta A_{||}$ for ITG at $\beta_e = 0.0022$ and KBM at $\beta_e = 0.02$ in figures \ref{itg_ms} and \ref{kbm_ms} (a)-(d). Both ITG and KBM show ballooning structure in $\delta\phi$ and anti-ballooning structure in $\delta A_{||}$, the ITG perturbation exhibits a more obvious ballooning angle than KBM which deviates from out-midplane, and KBM eigenmode structure is close to the ideal ballooning mode with self-adjointness in the high $\beta_e$ regime. Figure \ref{itg_ms} (e) shows the ITG polarization, where the amplitude of electrostatic parallel electric field $E_{||}^{ES} = -\mathbf{b_0}\cdot\nabla\delta\phi$ is almost the same with the net parallel electric field $E_{||}^{Net}= -\mathbf{b_0}\cdot\nabla\delta\phi-\left(1/c\right)\partial_t\delta A_{||}$, which indicates the quasi-electrostatic nature of ITG mode. In contrast, the $E_{||}^{Net}$ amplitude is much smaller than $E_{||}^{ES}$ for KBM due to the partial cancellation between longitudinal and transverse electric fields as shown in figure \ref{kbm_ms} (e), which indicates the KBM polarization is predominantly Alfv\'enic, i.e., $|E_{||}^{Net}|\sim 0\ll |E_{||}^{ES}|$. Building on above linear verifications of ITG and KBM, MAS can be applied for studying ion-scale drift wave instabilities in both electrostatic and electromagnetic regimes.


\section{Applications: bulk plasma kinetic effects on Alfv\'en eigenmodes in DIII-D geometry}\label{5}

Both theory \cite{berk1992, rosenbluth1992, mett1992} and experiment \cite{fasoli1995} have shown that the ideal MHD description of bulk plasmas is incomplete for AE activities, of which kinetic effects can damp AEs and thus determine the EP-driven thresholds, resolve the singularity arising from the resonance between MHD modes and Alfv\'en continua, and form kinetic AEs by trapping KAWs in the local potential well of Alfv\'en continuous spectra, etc. In this section, we carry out MAS simulation of AE activities of interest in DIII-D experimental geometry, including RSAE, TAE, KBAE and {\color{black}beta-induced Alfv\'en-acoustic eigenmode (BAAE)}, in which both the MHD and kinetic physics are well validated. The parallel and perpendicular heat ratios for Landau-fluid, full-MHD and reduced-MHD models are $\left(\Gamma_{i||},\Gamma_{i\perp}\right) = \left(3,7/4\right)$, $\left(\Gamma_{i||},\Gamma_{i\perp}\right)=\left(7/4,7/4\right)$ and $\left(\Gamma_{i||},\Gamma_{i\perp}\right) = \left(0,7/4\right)$, respectively.

\begin{figure}[H]
	\center
	\includegraphics[width=1\textwidth]{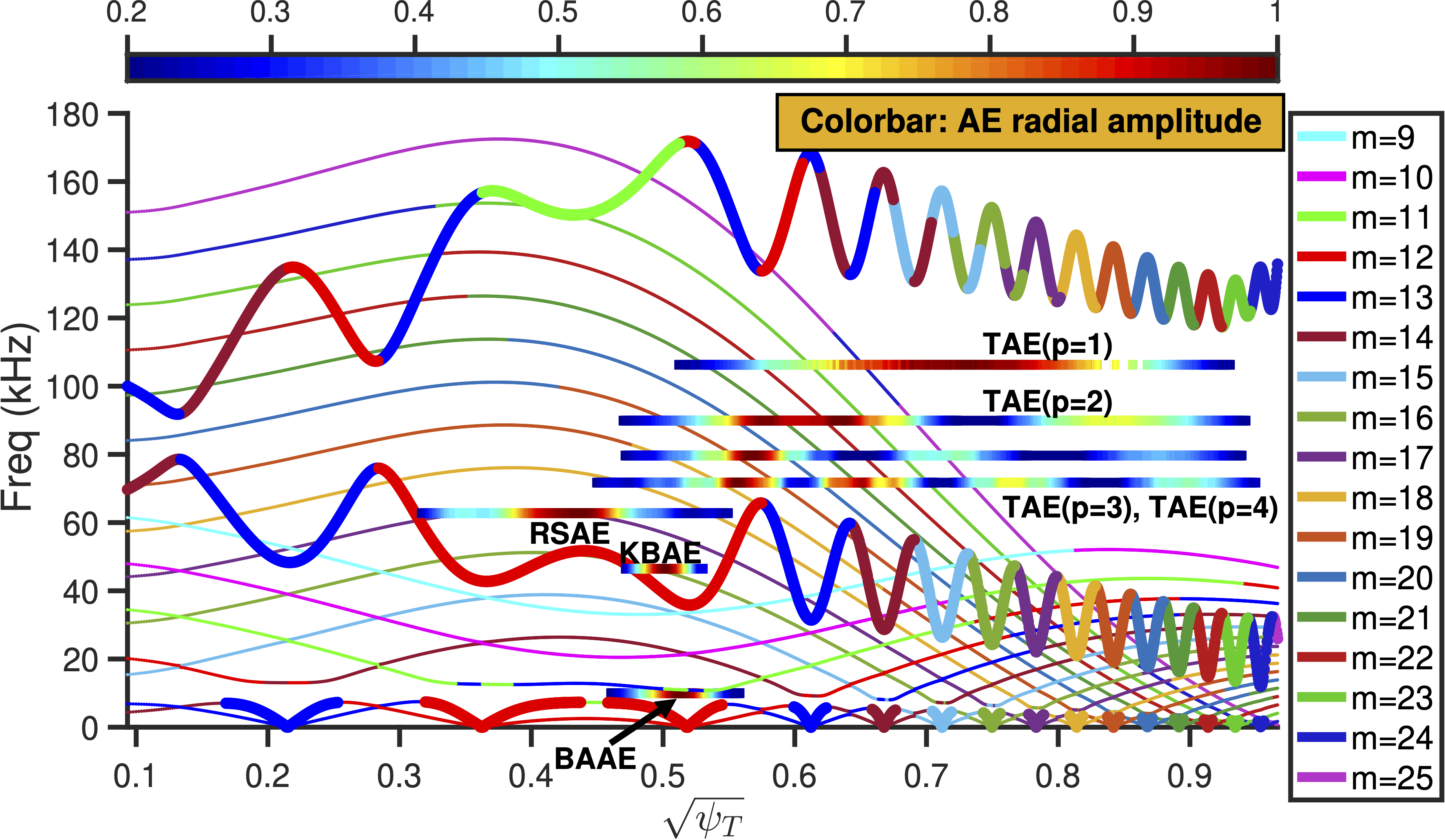}
	\caption{$n=4$ Alfv\'enic and acoustic continua for DIII-D shot \#159243. The thick and thin lines represent Alfv\'enic and acoustic branches respectively. The colorbar represents the normalized AE radial amplitude.}
	\label{full_continua}	
\end{figure}

We focus on DIII-D shot \#159243 with comprehensive diagnostics on RSAE and TAE \cite{cami2016}, of which the equilibrium at 805ms has been used for recent V\&V studies between gyrokinetic and kinetic-MHD hybrid codes \cite{sam2019}. The magnetic field geometry, safety factor $q$, bulk plasma density and temperature profiles ($n_e$, $T_e$, $T_i$) are described in figure 3 of Ref. \cite{sam2019}. The toroidal mode $n=4$ continuous spectra of SAW and ISW as well as discrete AEs are shown in figure \ref{full_continua}.

\subsection{RSAE: kinetic effects resolve MHD singularity arising from Alfv\'en continuum resonance}
RSAE is frequently observed in the NBI heated plasmas in the presence of a reversed magnetic shear, and the formation mechanism has been well understood theoretically \cite{Berk2001,Breizman2003,Breizman2005,fu2006}. At 805ms of DIII-D shot \#159243, RSAE is found as the dominant instability in recent V\&V study \cite{sam2019}, and is simulated by MAS here to clarify the roles of MHD and kinetic effects on RSAE physics.  We compute the RSAE mode structure and dispersion relation using hierarchical physics models introduced in sections \ref{2} and \ref{3}, specifically, the ideal reduced-MHD model using slow sound approximation in section \ref{2.3}, the ideal full-MHD model in section \ref{2.2}, and the Landau-fluid model without and with ion FLR terms labeled by \{Ion-FLR\} in section \ref{2} are used for comparison in order to delineate different level physics.
\begin{figure}[H]
	\center
	\includegraphics[width=1\textwidth]{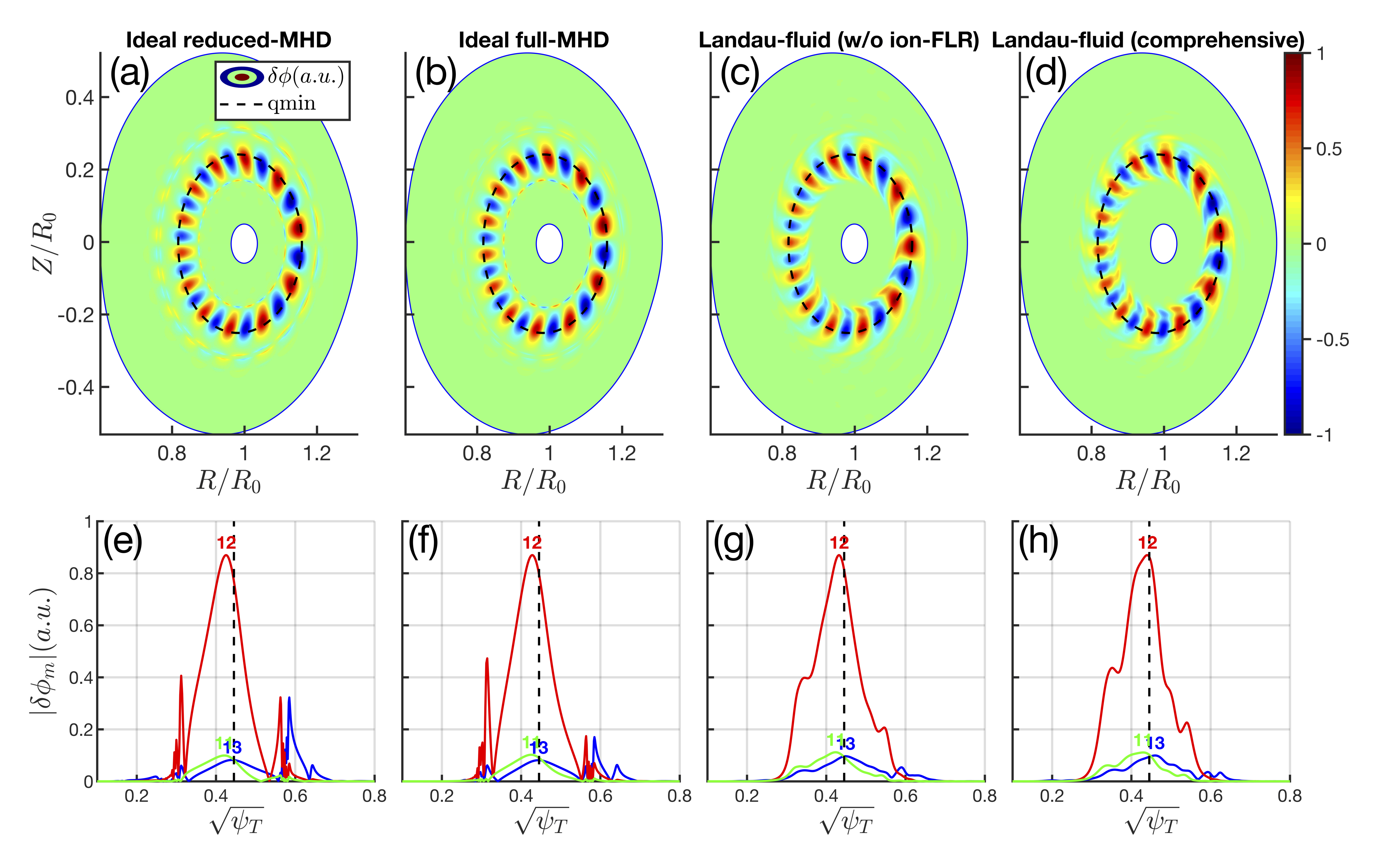}
	\caption{The 2D poloidal mode structure of $\delta\phi$ for n=4 RSAE using (a) ideal reduced-MHD model (f= 60.0kHz), (b) ideal full-MHD model (f= 60.6kHz), (c) Landau-fluid model without ion-FLR terms only ($f= 62.5kHz$, $\gamma_d/\omega_r = -1.82\%$), and (d) comprehensive Landau-fluid model (f= 62.8kHz, $\gamma_d/\omega_r = -1.67\%$). (e)-(h) are the corresponding radial structures of each poloidal harmonics.}
	\label{rsae}	
\end{figure}
The comparison of $n=4$ RSAE mode structure between four models in different levels are shown in figure \ref{rsae}. In the lowest order of MHD physics, all models exhibit that RSAE locates around $q_{min}$ = 2.945 surface, indicating that MAS can correctly deal with the basic nature of RSAE. Beyond the MHD physics, we shall show that the bulk plasma kinetic effects have non-trivial influences on the mode structure in the higher order. At the radial locations where RSAE resonates with Alfv\'en continuum (i.e., the intersections between RSAE and Alfv\'en continua in figure \ref{full_continua}), the reduced- and full-MHD models show unphysical spikes on the mode structure in figures \ref{rsae} (e) and (f), which is due to the MHD singularity arising from Alfv\'en resonance. In early MHD study, MHD singularity is avoided by excluding the Alfv\'en resonance region from the simulation domain when it does not affect the AE potential well much \cite{hu2014}, however, this method is not valid for our RSAE case. When the kinetic terms being responsible for small scale physics are retained in the model such as finite $E_{||}$, Landau damping and ion FLR effects etc., this MHD singularity can be resolved in a physical manner through SAW-KAW mode conversion in the resonance region, where the mode structures become smooth as shown in figures \ref{rsae} (g) and (h). Moreover, the KAWs suffer Landau damping and the residual parts (fine scale perturbations) superpose on the RSAE in the radial domain above the $m=12$ continuum where KAW can propagate, and it is also found that the ion-FLR effect can enhance KAW perturbations by comparing figures \ref{rsae} (g) and (h) which is in consistence with former kinetic study \cite{yu2009}. Meanwhile, the bulk plasma kinetic effects non-perturbatively modify the RSAE 2D poloidal mode structure by breaking the radial symmetry as shown in figures \ref{rsae} (c) and (d), which is characterized with a triangle shape with radial phase variation. On the other hand, the RSAE real frequencies from four models in figure \ref{rsae} are 60.0kHz, 60.6kHz, 62.5kHz and 62.8kHz respectively, and the kinetic influences are weak. However, the one-fluid MHD calculations cannot provide the damping rate information, while the Landau-fluid calculations give the damping rates $\gamma_d/\omega = -1.82\%$ and $\gamma_d/\omega = -1.67\%$ respectively by incorporating radiative damping, Landau damping and continuum damping, and the ion-FLR effect on damping rate is modest in consistence with the minor change of mode structure. The $n=4$ RSAE mode structures obtained from MHD and Landau-fluid models in figure \ref{rsae} are consistent with other perturbative and non-perturbative codes from figures 5 and 6 of Ref. \cite{sam2019} respectively, which indicate the reliability of MAS hierarchy physics model and the accuracy of numerical implementation.
\begin{figure}[H]
	\center
	\includegraphics[width=0.5\textwidth]{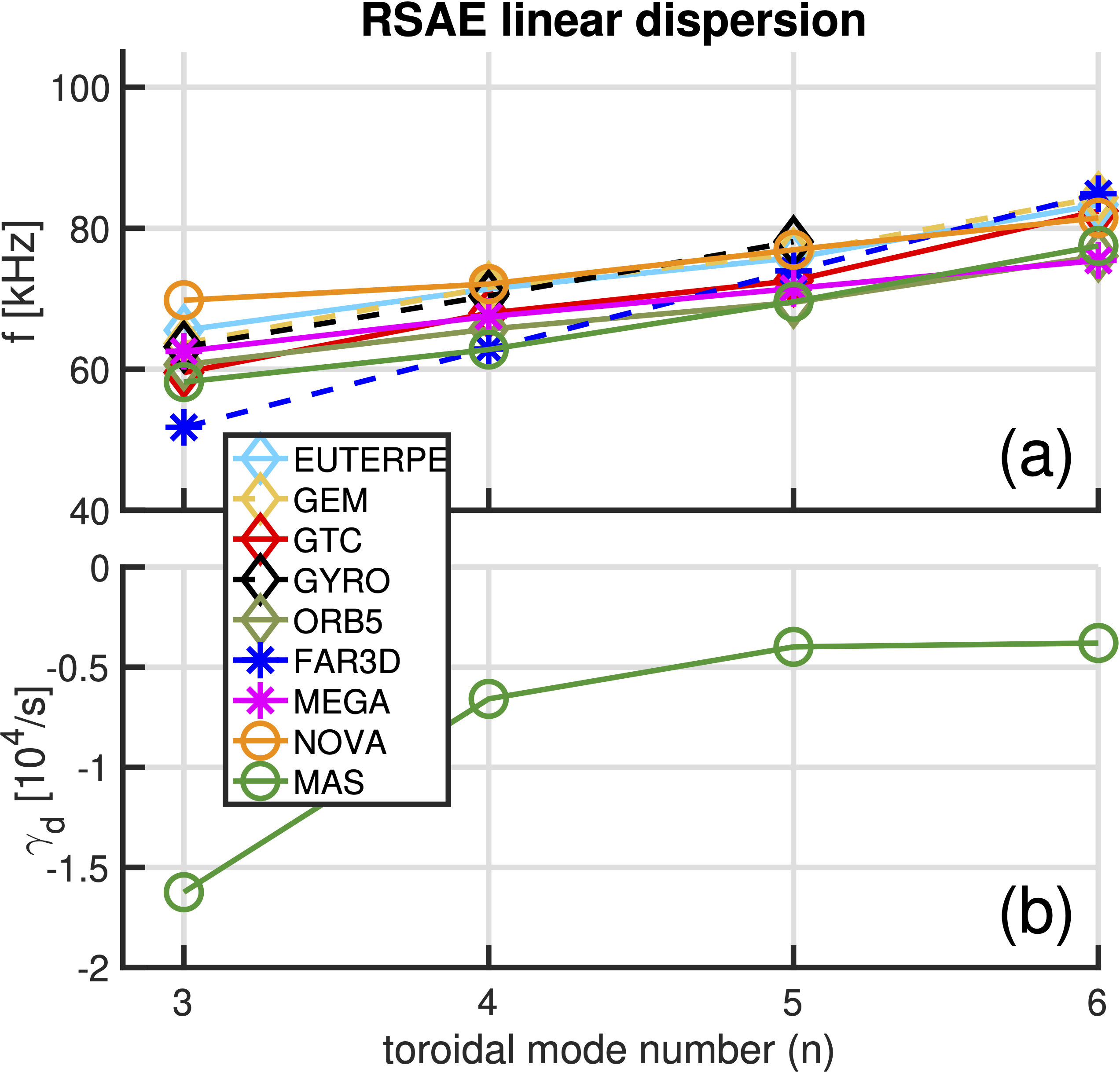}
	\caption{(a) The comparison of RSAE real frequency with different $n$ numbers between MAS and other gyrokinetic and kinetic-MHD hybrid codes in Ref. \cite{sam2019}. (b) Corresponding RSAE damping rates obtained from MAS.}
	\label{rsae_nsca}	
\end{figure}
We then simulate RSAEs from $n=3$ to $n=6$ using the comprehensive Landau-fluid model in section \ref{2}, and compare with other gyrokinetic and kinetic-MHD hybrid codes in Ref. \cite{sam2019} (note that EP non-perturbative effect on RSAE real frequency in this shot has been proved small). The RSAE real frequencies from MAS show good agreements with other codes in figure \ref{rsae_nsca} (a), and MAS further gives the damping rates in figure \ref{rsae_nsca} (b), which are ignored by the kinetic-MHD hybrid codes and are difficult to be measured in the initial value gyrokinetic codes. The increase of RSAE frequency with $n$ number in figure \ref{rsae_nsca} can be explained by following estimation of RSAE frequency in low-$\beta$ plasmas:
\begin{flalign}\label{rsae_omega}
	\begin{split}
		\omega_{RSAE} \approx \left[\left(\frac{m}{q_{min}}-n\right)^2\frac{V_{A}^2}{R_0^2} + \omega_{geod}^2\right]^{1/2} +\delta\omega
	\end{split},
\end{flalign}
where $\omega_{geod} = \sqrt{2\Gamma P_0/\left(m_in_{i0}\right)}/R_0$ is the geodesic acoustic frequency, and $\delta\omega$ represents the deviation of the discrete eigenmode from the reverse shear extreme point of Alfv\'en continuum. Analogous to the zero-point energy in a quantum mechanical system, $\delta \omega$ is determined by toroidicity, fast ion, pressure gradient, etc., and has a modest correction on $\omega_{RSAE}$. It has been known that the principal dominant poloidal harmonic $m$ of RSAE satisfies $m-0.5<nq_{min}<m$ \cite{Breizman2005}, thus it is straightforward to show that $\omega_{RSAE}$ increases with $\left(n,m\right) = \left(3,9\right), \left(4,12\right), \left(5,15\right), \left(6,18\right)$ in terms of $q_{min} = 2.945$ based on Eq. \eqref{rsae_omega}. 
\begin{figure}[H]
	\center
	\includegraphics[width=1\textwidth]{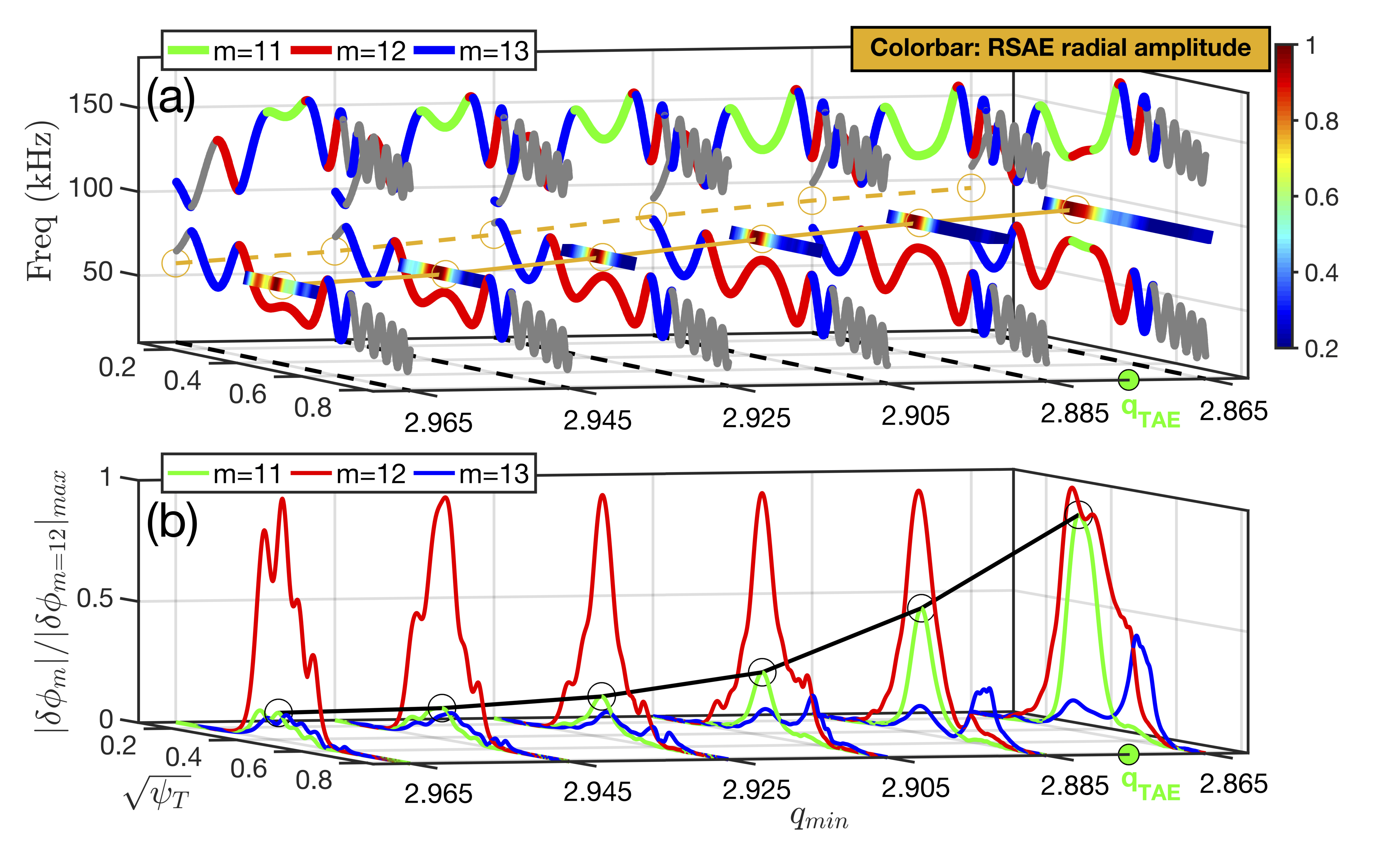}
	\caption{(a) $n=4$ Alfv\'en continua for various $q_{min}$ values. The colorbar represents the normalized RSAE radial amplitude and the yellow lines show the frequency dependence on $q_{min}$. (b) Corresponding radial profiles of poloidal harmonics $\delta \phi_m$. The black line shows the amplitude evolution of sub-dominant $m=11$ harmonic. The $q_{min}$ threshold (i.e., $q_{TAE}$) of RSAE-TAE transition is shown by the green circle.}
	\label{rsae_sweeping}	
\end{figure}
More importantly, Eq. \eqref{rsae_omega} indicates that $\omega_{RSAE}$ relies on $q_{min}$ sensitively, which commonly exhibits an upward frequency sweeping pattern when $q_{min}$ decreases with time from experimental observations \cite{cami2016,Berk2001,Breizman2003}. In order to further validate our model, we scan the $q_{min}$ in the range around the experimental value by adding a small constant $\Delta q$ to the q-profile in the simulation, and $m>nq_{min}$ is satisfied for all $q_{min}$ values with $n=4$ and $m=12$. It should be pointed out that the Grad-Shafranov relation breaks when finite $\Delta q$ is induced without making other modifications on equilibrium, however, this inconsistency is ignorable since $\Delta q/q_{min}\sim O(10^{-2})$. The $n=4$ Alfv\'en continua and RSAE amplitudes for the continuously varied $q_{min}$ are shown in figure \ref{rsae_sweeping} (a). As $q_{min}$ decreases, the extreme point of $m=12$ continuum upshifts and induces the RSAE to chirp up in frequency as shown by the yellow circle lines, and meanwhile the extreme point of $m=11$ continuum downshifts closely to $m=12$ continuum. When $q_{min}$ decreases below the RSAE-TAE transition threshold $q_{TAE} = \left(2m-1\right)/\left(2n\right) = 2.875$, the $m=11$ and $m=12$ continua begin to merge with each other by exchanging the extreme points, and two TAE gaps form at both sides of $q_{min}$ surface as shown by the $q_{min} =2.865$ case ($q_{min}<q_{TAE}$). The corresponding poloidal harmonic structures are shown in figure \ref{rsae_sweeping} (b), and it can be seen that as $q_{min}$ drops, the amplitude of sub-dominant $m=11$ harmonic becomes larger and finally comparable to the principal dominant $m=12$ harmonic, which indicates the transition from RSAE to TAE. So far, the main RSAE physics associated with bulk plasmas have been well validated with important kinetic characteristics in MAS, including mode structure/dispersion relation, RSAE-Alfv\'en continuum interaction/mode conversion, and RSAE-TAE transition.

\subsection{TAE: kinetic effects enable tunneling interaction between Alfv\'en gap mode and continuum}
TAE is also routinely observed in fusion experiments, which is formed by toroidicity and magnetic shear in the continuum gap of neighbouring poloidal harmonics $m$ and $m+1$. In the same DIII-D shot, the ECE data shows $n=4-6$ TAEs at the same time with RSAEs \cite{cami2016}. The TAE simulations in this subsection are performed using the comprehensive Landau-fluid model in section \ref{2}.
\begin{figure}[H]
	\center
	\includegraphics[width=1\textwidth]{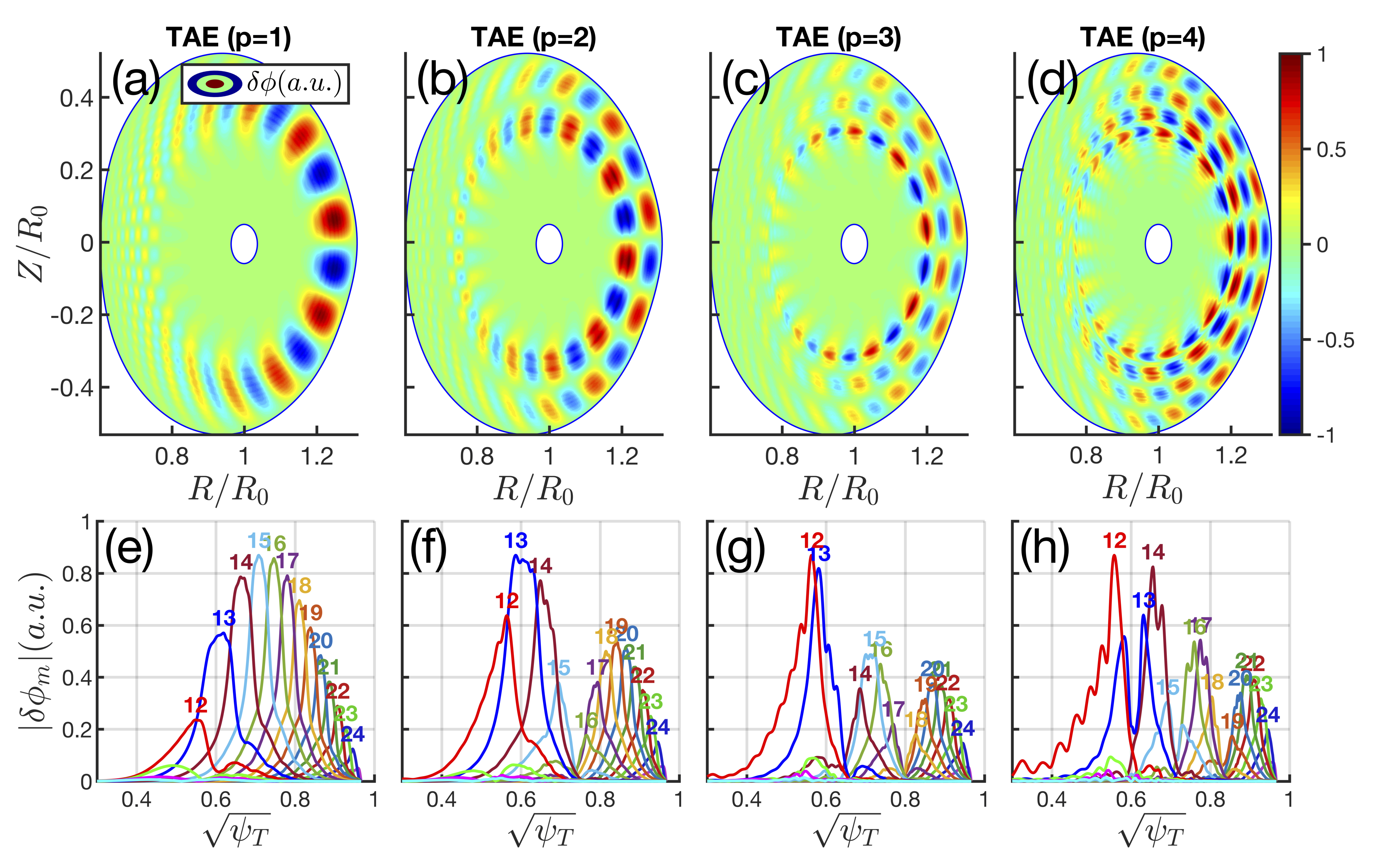}
	\caption{The 2D poloidal mode structure of $\delta\phi$ for $n=4$ TAE with different radial mode numbers. (a) $p=1$ with $f_{TAE} = 106.3kHz$ and $\gamma_d/\omega_r=-1.07\%$. (b) $p=2$ with $f_{TAE} = 89.9kHz$ and $\gamma_d/\omega_r = -1.1\%$. (c) $p=3$ with $f_{TAE} = 79.7kHz$ and $\gamma_d/\omega_r = -2.04\%$. (d) $p=4$ with $f_{TAE} = 71.7kHz$ and $\gamma_d/\omega_r = -3.01\%$.}
	\label{tae}	
\end{figure}
We first analyze the $n=4$ TAE mode structure, frequency and damping rate as shown in figure \ref{tae}. In DIII-D plasmas, the TAE exhibits radially global structure that consists of multiple poloidal harmonics. For $n=4$ mode, there are multiple TAE eigenstates with discrete frequencies as shown in figure \ref{full_continua}, and the corresponding mode structures are featured by radial quantum number $p$ varying from 1 to 4 as shown in figure \ref{tae}. As $p$ number increases, the TAE real frequency ($\omega_r$) decreases and the damping rate ($\gamma_d$) increases, and the ratios are $\gamma_d/\omega_r = -1.07\%, -1.1\%, -2.04\%, -3.01\%$ for $p=1,2,3,4$ respectively. It has been studied analytically \cite{mett1992} and numerically \cite{fu2005,lauber2005} that the bulk plasmas kinetic effects can lead to the tunneling interaction between TAE and Alfv\'en continuum in the absence of AE-continuum resonance, which still results in a superposition of KAW and TAE mode structures. In figures \ref{tae} (e)-(h), the fine scale structures correspond to KAW perturbations as the tunneling strength between TAE and Alfv\'en continuum, which increases with $p$ number and is determined by the higher order kinetic terms as well as the tunneling distance between TAE and the nearest SAW continuum.
\begin{figure}[H]
	\center
	\includegraphics[width=0.9\textwidth]{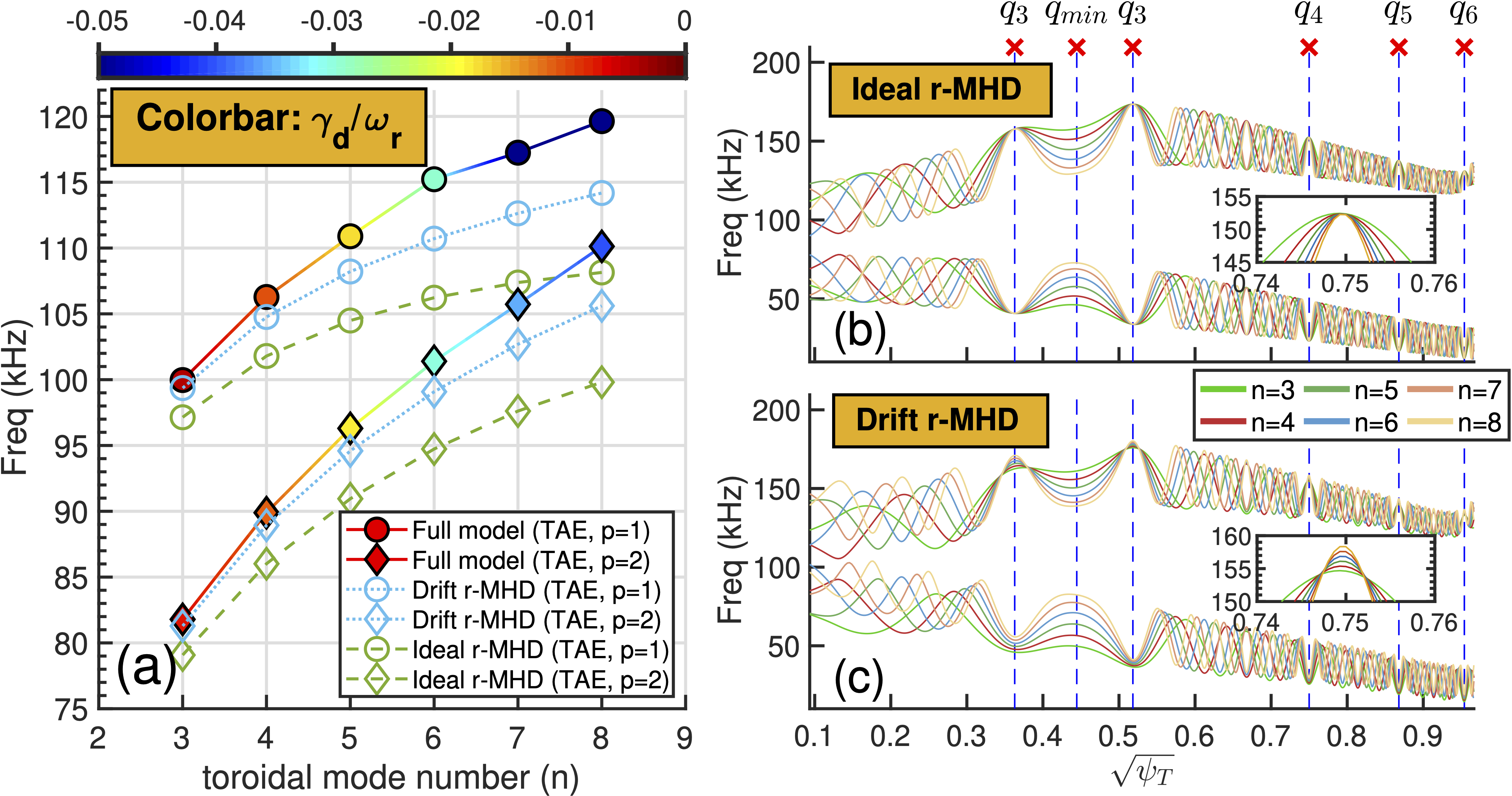}
	\caption{(a) The dependences of $p=1$ and $p=2$ TAE frequency and damping rate on $n$ number using ideal reduced-MHD, drift reduced-MHD and comprehensive Landau-fluid models. (b) Ideal reduced-MHD model and (c) drift reduced-MHD model calculations of Alfv\'en continua for different $n$ numbers.}
	\label{tae_damping}	
\end{figure}
Next, we choose the $p=1$ and $p=2$ TAEs for the $n$ number scan, which are weakly damped and observed in experiment \cite{cami2016,sam2019}. In order to clarify various physics effects on TAE dispersion relation, the ideal reduced-MHD, drift reduced-MHD, and comprehensive Landau-fluid models are applied for comparison in figure \ref{tae_damping}. All models indicate that both $p=1$ and $p=2$ TAE frequencies increase with $n$ number, while the detailed characteristics and the interpretations of underlying physics are listed below:
\begin{enumerate}
	\item In figure \ref{tae_damping} (a), the ideal reduced-MHD results indicate that the frequency of $p=1$ TAE branch approaches to a constant as $n$ number increases, and this constant corresponds to the high-$n$ ballooning mode theory prediction, which requires the separation between the radial scale lengths of plasma equilibrium, TAE envelope and poloidal harmonics.
	
	\item The ideal reduced-MHD Alfv\'en continua with different $n$ numbers are compared in figure \ref{tae_damping} (b), which show the same TAE gap structure. Thus the increase of TAE frequency with $n$ number is completely due to the breaking of scale-length separation condition in the ideal reduced-MHD framework.
	
	\item The increase of $p=2$ TAE frequency with $n$ number is faster than $p=1$ TAE branch as shown in figure \ref{tae_damping} (a). The $p=2$ TAE envelope scale-length is shorter than $p=1$ TAE, and it requires narrower poloidal harmonics (corresponding to higher $m$ and $n$ numbers) in order to satisfy the scale-length separation condition and approach the frequency limit predicted by the ballooning mode theory.
	
	\item The drift reduced-MHD simulation of TAE shows a higher frequency compared to the ideal reduced-MHD simulation for each $n$ number as shown in figure \ref{tae_damping} (a), and the frequency increase is close to the Alfv\'en continuum upshift by ion diamagnetic frequency $\omega_{p,i}^*$ measured from figure \ref{tae_damping} (c), thus the TAE frequency modification by plasma diamagnetic effect is mainly due to the change of TAE gap structure.
	
	\item The comprehensive Landau-fluid simulation shows the highest TAE frequency compared to the ideal and drift reduced-MHD results in figure \ref{tae_damping} (a), and the frequency differences come from the plasma compressibility and kinetic effects. As $n$ number increases, both $p=1$ and $p=2$ TAE frequencies get closer to the upper continuum of TAE gap, and consequently enhance the radiative damping.
\end{enumerate}
\begin{figure}[H]
	\center
	\includegraphics[width=1\textwidth]{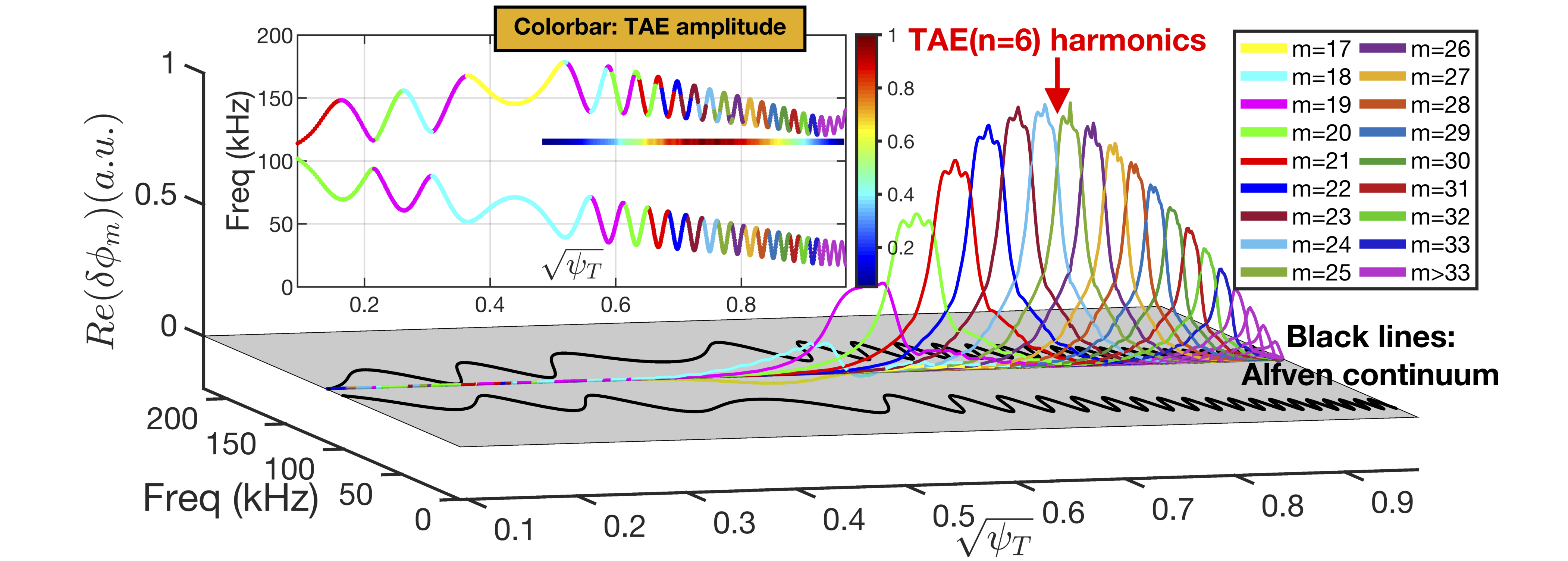}
	\caption{The interaction between TAE ($n=6$, $p=1$) and Alfv\'en continua. The radial structure of TAE poloidal harmonics is computed using comprehensive Landau-fluid model, and Alfv\'en continua is computed using drift reduced-MHD model with leading diamagnetic effect.}
	\label{tae_n6}	
\end{figure}
\begin{figure}[H]
	\center
	\includegraphics[width=1\textwidth]{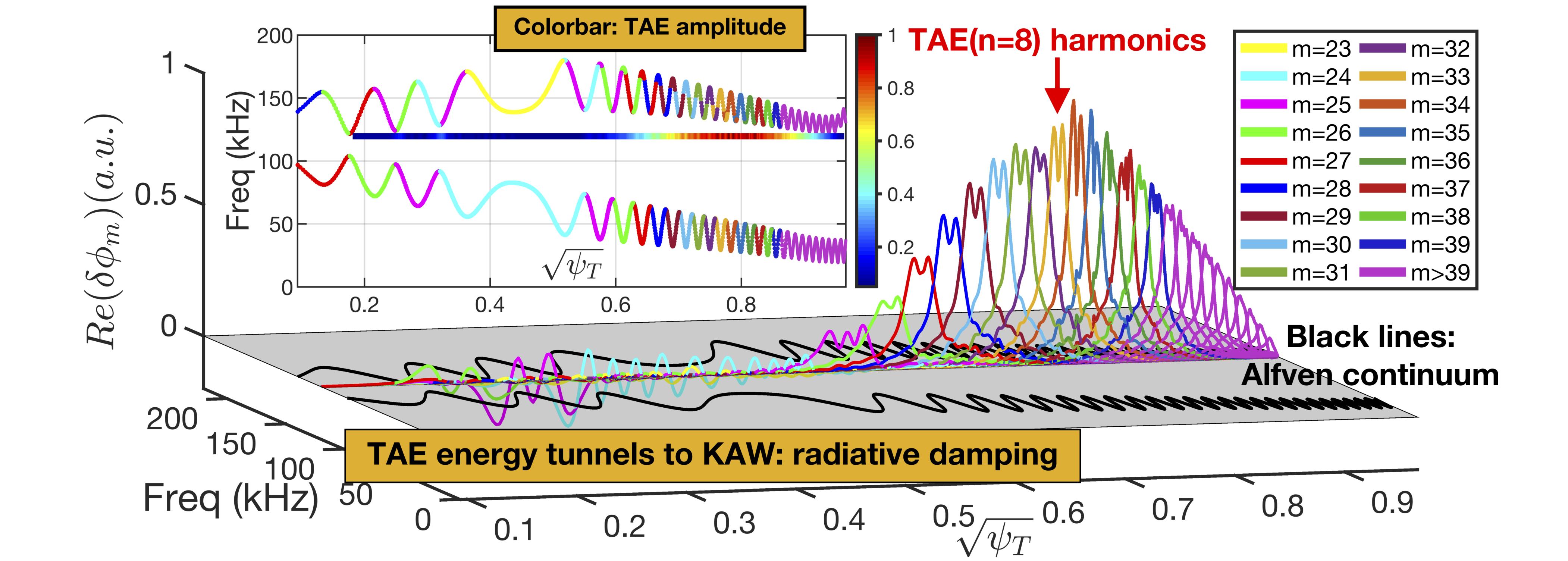}
	\caption{The interaction between TAE ($n=8$, $p=1$) and Alfv\'en continua. Other details are the same with figure \ref{tae_n6}.}
	\label{tae_n8}	
\end{figure}

In order to further elucidate the tunneling mechanism between TAE gap mode and Alfv\'en continuum, we compare the $n=6$ and $n=8$ TAE mode structures with the same radial quantum number $p=1$ based on the Landau-fluid simulations as shown in figures \ref{tae_n6} and \ref{tae_n8}, where the TAE frequencies, radial positions and amplitude contours are plotted together with the Alfv\'en continua. For $n=6$ case in figure \ref{tae_n6}, the radial profile of each poloidal harmonic exhibits the fine scale perturbation with small amplitude on the top, because the KAWs are excited at TAE frequency through tunneling effect in presence of the fourth-order kinetic terms such as the ion-FLR and finite $E_{||}$. Here, the KAW frequency and $n$ number are the same with TAE, and the coupling location associated with $k_{||}$ is determined by the KAW dispersion relation. For comparison, the tunneling effect in $n = 8$ case is much stronger than $n=6$ case, along with the higher amplitude of fine scale structure in figure \ref{tae_n8} as well as the larger damping rate in figure \ref{tae_damping} (a), and the main causes can be summarized as  
\begin{enumerate}
	\item $n=8$ TAE frequency is more closer to the accumulating points of upper continuous spectra, which shortens the tunneling distance and then reduces the cut-off loss of KAW energy.
	
	\item The radial width of TAE poloidal harmonic of $n=8$ case is narrower than $n=6$ case, which gives rise to a higher $k_\perp$ value and increases the kinetic terms such as ion-FLR and finite $E_{||}$ for KAW excitation.
	
	\item  The amplitude of KAW perturbation is related to the radiative damping strength, namely, a fraction of TAE energy tunnels to the short-wavelength structure so that the kinetic dissipations (such as Landau damping) can take place.
\end{enumerate}

In addition, the $n=8$ TAE-KAW coupling extends to the plasma core region (in the range of $0.2<\sqrt{\psi_T}<0.35$) with large amplitude oscillations in figure \ref{tae_n8}, which no longer attributes to the tunneling effect. Instead, the TAE frequency touches the upper continuous spectra in the core region and Alfv\'en resonance happens, which results in the mode conversion and KAW excitation. In the plasma edge ($\sqrt{\psi_T} > 0.85$), although $m>39$ poloidal harmonics of $n=8$ TAE are close to the upper continuum, the KAWs due to the tunneling interaction or Alfv\'en resonance are not obvious, because the low edge temperatures weaken the kinetic effects.

\subsection{KBAE and BAAE: kinetic effects determine the existences of low-frequency Alfv\'enic fluctuations}
The low frequency Alfv\'en eigenmodes, KBAE with $\omega\sim \left[7T_i/\left(2T_e\right)+2 \right]^{1/2}C_s/R_0$ and BAAE with $\omega \sim C_s/\left(qR_0\right)$ in toroidal plasmas, are also studied using MAS Landau-fluid simulations. In DIII-D discharge \# 159243, KBAE and BAAE are not observed together with RSAE and TAE, since the bulk plasma damping becomes heavier in the low frequency regime which exceeds the EP drive and stabilizes the modes. Although EP drive is not incorporated in this work, MAS is able to analyze the stable KBAE and BAAE by providing the damping rate, frequency and mode structure in a non-perturbative manner, which could serve as important channels for indirect heating of bulk plasmas by energetic particles (include $\alpha$ particle). As shown in figure \ref{full_continua}, there exist both $n=4$ KBAE and BAAE at the rational surface of $q=3$ in the frequency regime being close to the theoretical prediction, and the details are explained in the following.

The BAE gap is generated due to the geodesic magnetic curvature assoicated with $E\times B$ drift compression in finite-$\beta$ plasmas, which is below the TAE gap as shown in figure \ref{full_continua}. In the continuous spectra between TAE and BAE gaps, the {\color{black}continuum accumulation point (CAP)} at the rational surface with $k_{||}\sim 0$ has a finite frequency, which is called BAE-CAP frequency as $\omega_{BAE-CAP} = \left[7T_i/\left(2T_e\right)+2 \right]^{1/2}C_s/R_0$. When the kinetic effects are incorporated, the continuous SAW spectra around BAE CAP are discretized into radially bounded states, i.e., KBAEs, which are caused by trapping the inward propagating KAWs in the continuum potential wells. In figure \ref{full_continua}, a typical KBAE locates in the continuum potential well of $m=12$ SAW, and the mode frequency and damping rate are $f_{kbae} = 46.5kHz$ and $\gamma_d/\omega_r = -6.1\%$, respectively, which suffers larger bulk plasma damping than RSAE. On the other hand, the center of BAAE gap is around 10 kHz in figure \ref{full_continua}, which is formed due to the coupling of Alfv\'enic and acoustic continua through geodesic magnetic curvature. Here, the criterion of  Alfv\'enic and acoustic polarizations for the ideal full-MHD continua in figure \ref{full_continua} is determined by the maximal amplitude poloidal harmonic among the coupled SAWs and ISWs. It is seen that the upper continuous spectra of BAAE gap only contain purely acoustic branch while the lower ones contain both Alfv\'enic and acoustic branches, which is consistent with former theory and simulation \cite{Gorelenkov2007a, zhang2016}. In the BAAE gap, a discretized BAAE is found above the $m=12$ Alfv\'en continuum with a low frequency of $f_{baae} = 10kHz$ and a large damping rate of $\gamma_d/\omega_r = -25.1\%$. {\color{black}Regarding to the experimental relevance, the low-frequency Alfv\'enic fluctuations in the BAAE gap of Alfv\'en continuum have been observed during sawtooth cycle in ASDEX Upgrade \cite{curran2012}, DIII-D \cite{heidbrink2020b} and EAST \cite{xu2020} etc, of which frequency increases with increasing $\nabla T_e$, however, it has been debated for a long time that whether the heavily damped BAAE really exist, and recent gyrokinetic analyses support the view that the reactive-type KBM instability with $\omega\sim\omega_{*p,i}$ is responsible for these low-frequency Alfv\'en modes \cite{Lauber2013, curran2012, ma2022}.}

\begin{figure}[H]
	\center
	\includegraphics[width=1\textwidth]{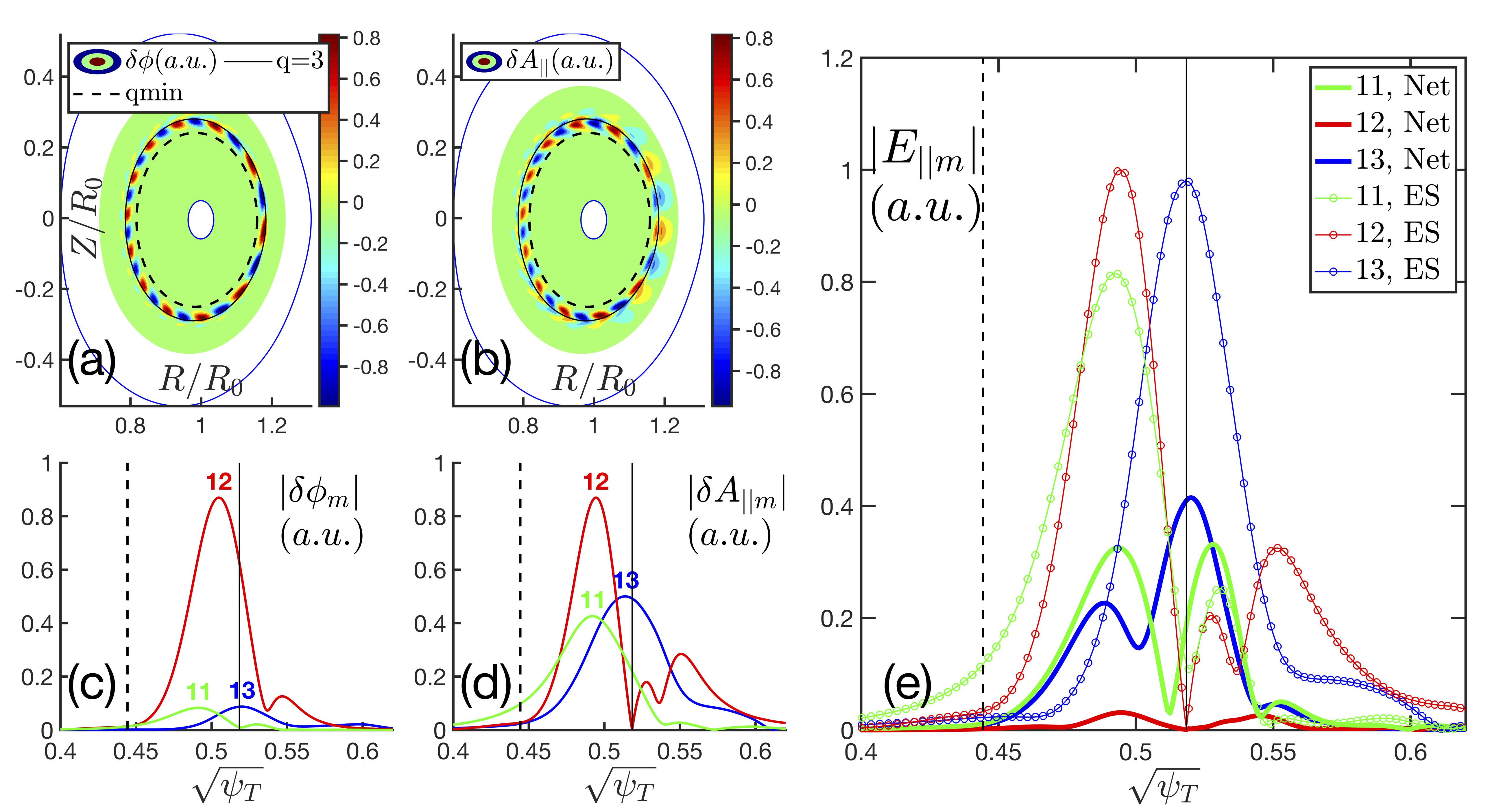}
	\caption{The n=4 KBAE mode structures. (a)-(d): The poloidal contour plots and poloidal harmonic radial plots of electrostatic potential $\delta \phi$ and parallel vector potential $\delta A_{||}$. (e) The poloidal harmonic radial plots of parallel electric field $E_{||}$, where the thin lines with circle represent $E_{||}^{ES}=-\mathbf{b_0}\cdot\nabla\delta\phi$ and the thick lines represent $E_{||}^{Net}= -\mathbf{b_0}\cdot\nabla\delta\phi-\left(1/c\right)\partial_t \delta A_{||}$.}
	\label{bae}	
\end{figure}

\begin{figure}[H]
	\center
	\includegraphics[width=1\textwidth]{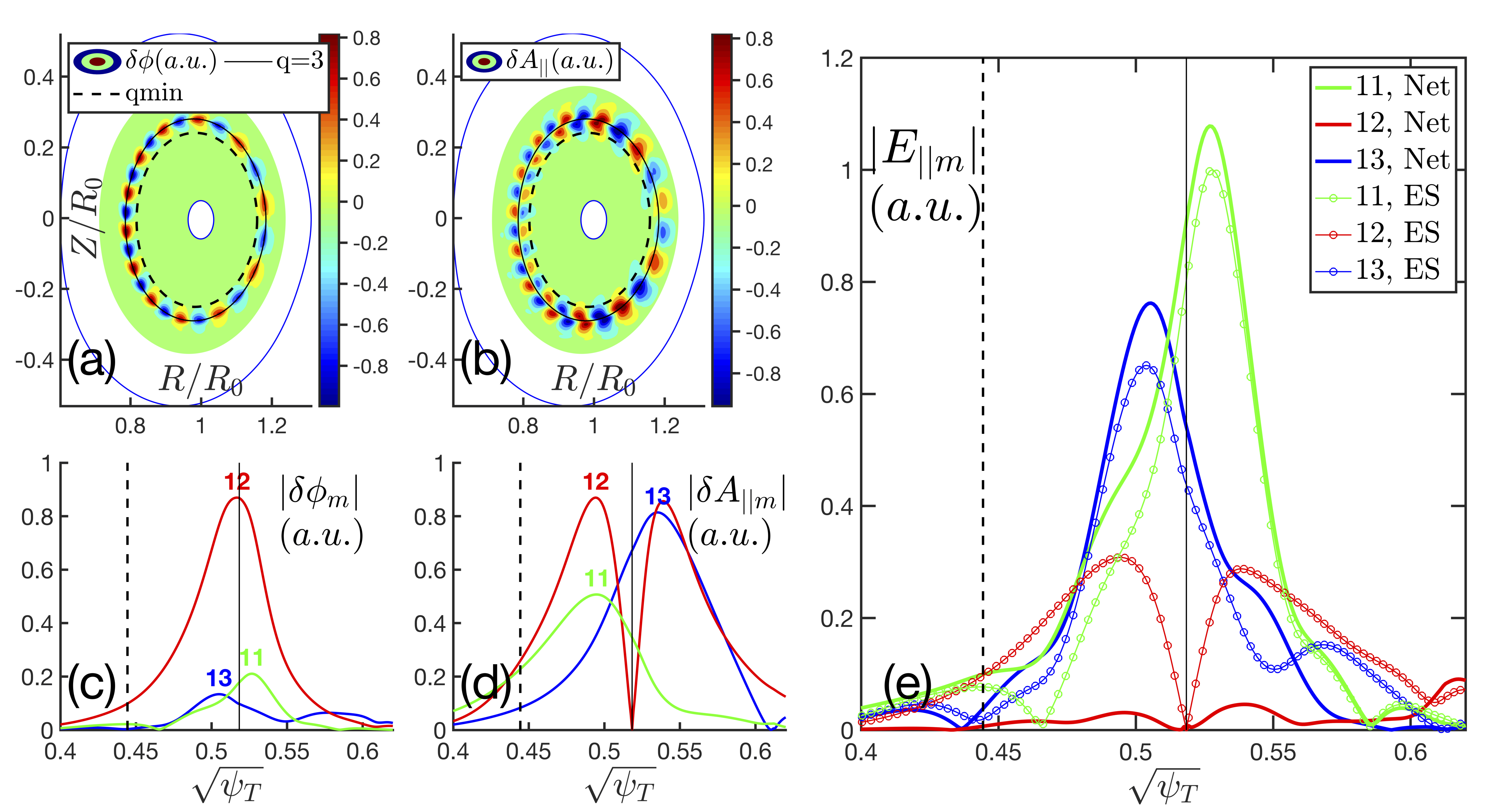}
	\caption{The n=4 BAAE mode structures. The captions of (a)-(e) are the same with figure \ref{bae}.}
	\label{baae3}	
\end{figure}

The mode structures and polarizations of KBAE and BAAE are shown in figures \ref{bae} and \ref{baae3}, respectively. Both the KBAE and BAAE exhibit weakly ballooning structures in the electrostatic potential $\delta\phi$, which are dominated by the $m =12$ principal poloidal harmonic and $m=11$ and $m=13$ poloidal sidebands with smaller amplitudes. From figure \ref{bae} (c), it is noted that the peak amplitude of $m=12$ harmonic of KBAE deviates from the q = 3 rational surface where $k_{||}\sim 0$, which is due to the radial symmetry breaking of the $m=12$ continuum potential well above the BAE-CAP frequency as shown in figure \ref{full_continua}. For comparison, the peak amplitude of $m=12$ harmonic in BAAE is located at q = 3 rational surface (i.e., Alfv\'enic-acoustic coupling location) as shown in figure \ref{baae3} (c), which is consistent to the results of theory and gyrokinetic simulation in presence of normal magnetic shear. Compared to the $\delta \phi$ structure, the amplitudes of $m=11$ and $m=13$ poloidal sidebands of $\delta A_{||}$ increase and get close to the $m=12$ principal poloidal harmonic for both KBAE and BAAE, which are shown in figures \ref{bae} (c) and (d) and figures \ref{baae3} (c) and (d). Moreover, the radial structures of electrostatic parallel electric field $E_{||}^{ES} = -\mathbf{b_0}\cdot\nabla\delta\phi $ and net parallel electric field $E_{||}^{Net} = -\mathbf{b_0}\cdot\nabla\delta\phi - \left(1/c\right)\partial_t \delta A_{||}$ are shown in figures \ref{bae} (e) and \ref{baae3} (e) for KBAE and BAAE, respectively. One can see that $|E_{||}^{Net}| \ll |E_{||}^{ES}|$ is valid for all poloidal harmonics of KBAE in figure \ref{bae} (e), indicating the Alfv\'enic polarization that $\delta \phi$ and $\delta A_{||}$ terms in $E_{||}^{Net}$ are comparable and mainly cancel with each other. For BAAE case in figure \ref{baae3} (e), Alfv\'enic polarization (i.e., $|E_{||}^{Net}|\ll |E_{||}^{ES}|$) is only valid for the $m=12$ principal poloidal harmonic, while the $m=11$ and $m=13$ poloidal sidebands exhibit the acoustic polarization since $|E_{||}^{Net}| \sim |E_{||}^{ES}|$ and $\delta A_{||}$ effect is weak. This special polarization nature of BAAE is also discovered by the former MHD \cite{Gorelenkov2007a,Gorelenkov2007b} and gyrokinetic simulations \cite{zhang2016, liu2017}. 
Based on MAS simulations, we conclude that the bulk plasma kinetic effects are crucial to the existences of KBAE and BAAE:
\begin{enumerate}
	\item The formation of discretized KBAE requires the high order kinetic terms beyond the MHD model. Our Landau-fluid model covers KBAE physics by keeping the ion-FLR, diamagnetic drift and Landau damping effects being consistent with the gyrokinetic model \cite{zonca98,wang2010,ma2015,lu2018}.
	
	\item Previous gyrokinetic simulations \cite{zhang2016,liu2017} have shown that the BAAE in nomenclature \cite{Gorelenkov2007a} suffers severe Landau damping in the regime of $T_i\sim T_e$, which is beyond the one-fluid MHD physics. Our Landau-fluid model incorporates the Landau damping physics in BAAE simulation (recall $\gamma_d/\omega_r = -25.1\%$), and supports recent GFLDR theory analysis \cite{chen2017} on the view that BAAE is more difficult to be excited than other AEs.
\end{enumerate}

\section{Summary}\label{6}
In this work, a five-field Landau-fluid eigenvalue code MAS has been developed to analyze the bulk plasma stability with kinetic effects. The formulation of MAS physics model has been well examined and gauged through comparing with ideal MHD, electrostatic ion-fluid and drift-kinetic models in different limits. Meanwhile, MAS is numerically verified and validated for the typical reactive- and dissipative-type eigenmodes in fusion plasmas, including internal kink, ITG and KBM. Furthermore, the common AE activities in experiments, such as frequency sweeping of RSAE, radiative damping of TAE, and low frequency AEs including KBAE and BAAE, have been produced by MAS using the experimental parameters of a well-diagnosed DIII-D shot. The main physical and numerical features of MAS can be summarized as follows.
\begin{enumerate}
	\item \textbf{Broad range of validity regime.} The Landau-fluid physics model in MAS can recover ideal full-MHD, electrostatic ion-fluid and drift-kinetic models in the limits of long wavelength, electrostatic and uniform plasma, respectively.
	\item \textbf{Bulk plasma kinetic effects.} Besides the fundamental MHD physics, MAS Landau-fluid physics model further incorporates various kinetic effects of bulk plasma self-consistently, including ion and electron diamagnetic drifts, ion-FLR, ion and electron Landau resonances and finite parallel electric field $E_{||}$.  
	\item \textbf{Nonperturbative approach.} With retaining abundant kinetic effects, the Landau-fluid equation set still can be casted to a generalized linear eigenvalue problem for $\omega$, which does not involve the complex $\omega$ dependence and has the numerical convenience similar to MHD models \cite{kerner1998,xie2015}. Consequently the eigenmodes are solved nonperturbatively with respect to the kinetic terms, which can give the kinetic-determined and kinetic-modified mode structures directly.
	\item \textbf{Experimental geometry.} MAS code is developed based on the realistic experimental geometry using Boozer coordinates, which has simple forms of Jacobian and $\mathbf{B_0}$ field and thus simplifies the physics formulation expanded in complex geometry while still captures all geometry effects through the metric tensors.
	\item \textbf{Resolve MHD singularity.} The MHD singularity arising from the interactions between discretized eigenmodes and continuous spectra with either Alfv\'enic or acoustic polarization, is well-resolved in MAS through keeping the small-scale kinetic terms in Landau-fluid model, which results in the smooth mode structures at the interaction location rather than the spikes in traditional MHD calculation. Thus, MAS can properly address the bulk plasma damping effects on AEs in a physical manner, namely, continuum damping, Landau damping and radiative damping.
	\item \textbf{Wide and practical applications for fusion plasma problems.} MAS has been successfully benchmarked and validated for the Landau damping of normal modes (e.g., KAW and ISW), the macroscopic MHD modes (e.g., kink), the microscopic drift-wave instabilities in low- and high-$\beta$ regimes (e.g., ITG and KBM), and typical AEs (e.g., RSAE, TAE, KBAE and BAAE) with different polarizations which are widely observed in experiments.
	\item \textbf{Well-circulating limit for bulk plasma.} The moment equations in Landau-fluid model can well approximate the bulk plasma dynamics in the well-circulating limit, which is more accurate for high aspect-ratio device with a large fraction of passing particles. 
\end{enumerate}

We present the physics model, numerical scheme, verifications and validations of MAS code systematically, as the demonstration of an efficient tool for analyzing plasma activities in theoretical and experimental applications. {\color{black} MAS code and data in this work are available for collaborative communication, please contact the first two authors for more details.} Although MAS is capable of describing many ion-scale waves and instabilities with essential kinetic effects, it still lacks of the trapped particle effects of bulk plasma such as neoclassical inertia \cite{ma2022} and trapped electron mode (TEM) drive \cite{li2022}, which can be important in the regimes of low frequency (i.e., close to thermal ion magnetic drift frequency $\omega\sim\omega_{di}$) and low aspect-ratio. On the other hand, the EP resonance associated to magnetic drift is necessary for AE excitations, of which response function with important FLR and FOW effects cannot be easily fitted by Landau closure technique, thus in general EP dynamics require a gyrokinetic treatment with high fidelity \cite{zonca1996,lauber2007}. We plan to extend MAS formulation with including trapped thermal particle and EP dynamics in the future studies, and meanwhile keep the computational advantage of high efficiency with these upgrades.

\section{ACKNOWLEDGMENTS}
This work is supported by the National MCF Energy R\&D Program under Grant Nos. 2018YFE0304100
and 2017YFE0301300; National Natural Science Foundation of China under Grant Nos. 12275351, 11905290 and 11835016; and the start-up funding of Institute of Physics, Chinese Academy of Sciences under Grant No. Y9K5011R21. We would like to thank useful discussions with Liu Chen, Xiaogang Wang, Guoyong Fu, Zhixin Lu, Wei Chen, Zhiyong Qiu, Zhengxiong Wang, Jiquan Li, Huishan Cai, Zhibin Guo, Ruirui Ma and Guillaume Brochard. 

\appendix
\section{The orientation convention of basis vector in Boozer coordinates}\label{A5}
\begin{figure}[H]
	\center
	\includegraphics[width=1\textwidth]{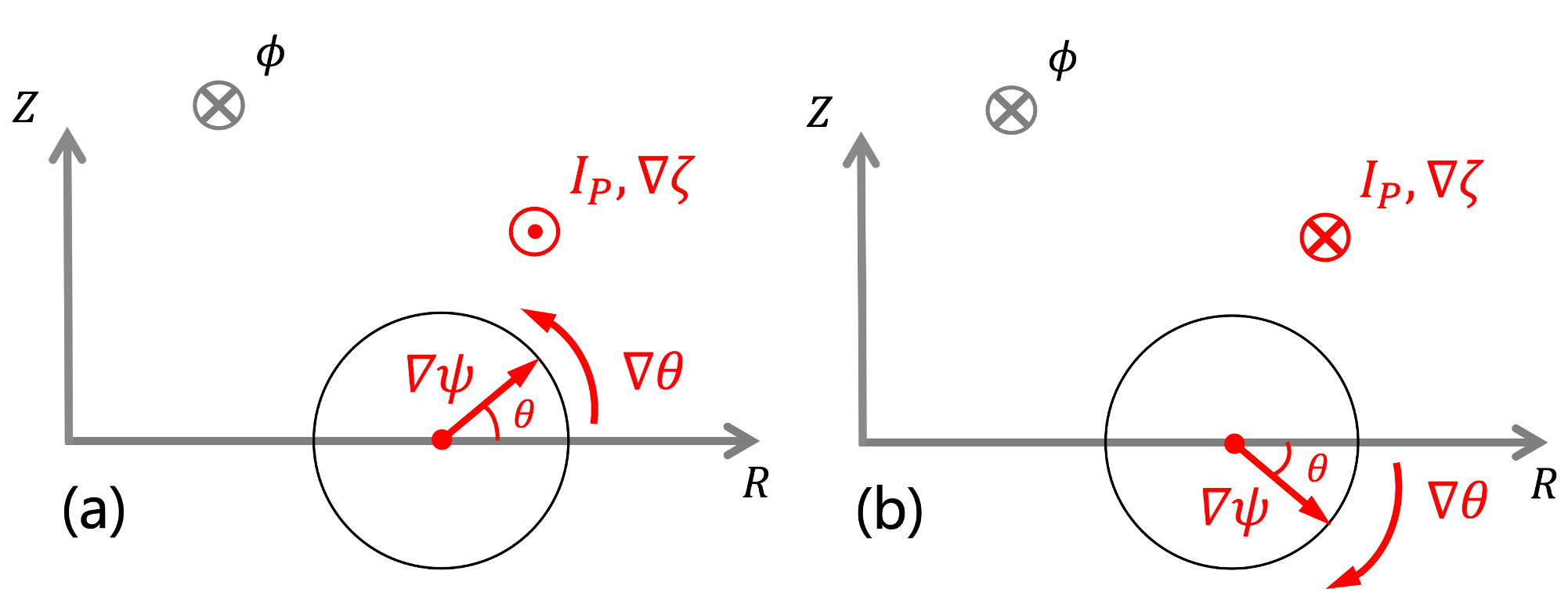}
	\caption{The reference right-handed cylindrical coordinates $\left(R,\phi,Z\right)$ in EFIT are shown by the grey color, where the unit vector $\boldsymbol{\hat{\phi}}$ is along the counter-clockwise direction from the top view of tokamak. (a) $I_p$ and $\nabla\zeta$ are opposite to the direction of $\boldsymbol{\hat{\phi}}$, and $\nabla\theta$ is along the counter-clockwise direction in the poloidal plane. (b) $I_p$ and $\nabla\zeta$ are along the direction of $\boldsymbol{\hat{\phi}}$, and $\nabla\theta$ is along the clockwise direction in the poloidal plane.}
	\label{basis_vector}	
\end{figure}

MAS uses the right-handed Boozer coordinates $\left(\psi,\theta,\zeta\right)$ of which contravariant basis vectors satisfy $\nabla\psi\times\nabla\theta\cdot\nabla\zeta = J^{-1} > 0$. Specifically, $\nabla\psi$ is orthogonal to the magnetic flux surface and always points to the edge, $\nabla\zeta$ is chosen to be along the co-current direction (same as the toroidal plasma current $I_p$ direction), and $\nabla\theta$ direction can then be determined by the right-handed rule. Each combination of contravariant basis vector directions is shown in figure \ref{basis_vector}, and the EFIT coordinates $\left(R,\phi,Z\right)$ (right-handed cylindrical coordinates) are shown as the frame of reference..

\section{Ideal full-MHD equations in $\left(\delta \phi, \delta A_{||}, \delta B_{||}\right)$ potentials}\label{A3}
The one-fluid MHD model using electromagnetic fields $\left(\mathbf{E},\mathbf{B}\right)$ has been widely applied to describe the bulk plasma dynamics in many fluid and fluid-kinetic hybrid codes. It is useful to map the classic ideal full-MHD equation set using $\left(\mathbf{E},\mathbf{B}\right)$ fields to the form using $\left(\delta\phi, \delta A_{||}, \delta B_{||}\right)$ potentials, which could connect our Landau-fluid model and ideal full-MHD model. Let us start from the linearized ideal full-MHD equation set in terms of $\left(\mathbf{E},\mathbf{B}\right)$
\begin{flalign}\label{momu_eq}
\begin{split}
\rho_0\frac{\partial \mathbf{\delta u}}{\partial t}
= -\nabla\delta P + \frac{1}{c}\mathbf{\delta J}\times \mathbf{B_0}+\frac{1}{c}\mathbf{J_0}\times\mathbf{\delta B},
\end{split}
\end{flalign}
\begin{flalign}\label{ampere}
\begin{split}
\nabla \times \mathbf{\delta B} = \frac{4\pi}{c}\mathbf{\delta J},
\end{split}
\end{flalign}
\begin{flalign}\label{Faraday}
\begin{split}
\frac{\partial \mathbf{\delta B}}{\partial t} = -c\nabla\times \mathbf{\delta  E}
\end{split}
\end{flalign}
\begin{flalign}\label{ohm}
\begin{split}
\mathbf{\delta E}+\frac{1}{c} \mathbf{\delta u}\times \mathbf{B_0}=0,
\end{split}
\end{flalign}
\begin{flalign}\label{deltaP}
\frac{\partial \delta P}{\partial t}
=-\mathbf{v}\cdot\nabla P_0 - \Gamma P_0\nabla\cdot\mathbf{v},
\end{flalign}
where $ \mathbf{\delta u}$ is the MHD perturbed velocity, $\rho_0 = n_{i0}m_i$ is the plasma mass density, $\Gamma$ is the one-fluid MHD heat ratio, $P_0 = P_{e0} + P_{i0}$ and $\delta P = \delta P_e + \delta P_i$ are the total equilibrium and perturbed pressures, $\mathbf{J_0}$ and $\mathbf{\delta J}$ are the equilibrium and perturbed currents, $\mathbf{B_0}$ and $\mathbf{\delta B}$ are the equilibrium and perturbed magnetic fields, and $\mathbf{\delta E}$ is the perturbed electric field. 

Using Eq. \eqref{ohm} and the definition $\mathbf{\delta E} = -\nabla\delta\phi - \left(1/c\right)\partial_t\mathbf{\delta A}$, we can derive the parallel Ohm's law straightforwardly, which serves as the dynamic equation for $\delta A_{||}$
\begin{align}\label{mhd-ohm}
	\frac{\partial \delta A_{||}}{\partial t} = -c\mathbf{b_0}\cdot\nabla\delta\phi.
\end{align}
Meanwhile, we can decompose the MHD velocity into the electrostatic and inductive parts of perpendicular $\mathbf{E}\times\mathbf{B}$ drift and the parallel motion
\begin{align}\label{v_ideal_MHD}
\mathbf{\delta u}
=\frac{c\mathbf{b_0}\times\nabla_\perp \delta\phi}{B_0} + \frac{1}{B_0}\mathbf{b_0}\times\frac{\partial \mathbf{\delta A_\perp}}{\partial t} + \delta u_{||}\mathbf{b_0}.
\end{align}

The unstable MHD modes in strongly magnetized plasmas are commonly characterized by the mode structures elongated along $\mathbf{B_0}$ field lines with $k_{||}\ll k_\perp$, then the approximations $\nabla \times \left(\delta A_{||}\mathbf{b_0}\right) \approx \mathbf{\delta B_\perp}$ and $\nabla\times\mathbf{\delta A_\perp}\approx \delta B_{||}\mathbf{b_0}$ are valid for most plasma regimes of interest. Following Refs. \cite{Breizman2005} and \cite{fu2006}, we can derive the coupled equations for $\delta\phi$, $\delta B_{||}$ and $\delta u_{||}$ by taking the proper projections of Eq. \eqref{momu_eq} with considering Eq. \eqref{v_ideal_MHD}. First, we operate $\nabla\cdot\left[\left(\boldsymbol{b_0}/B_0\right)\times\cdots\right]$ on both sides of Eq. \eqref{momu_eq}, and yield the ideal MHD vorticity equation as the dynamic equation for $\delta\phi$
\begin{flalign}\label{mhd-vorticity}
\begin{split}
\frac{\partial }{\partial t}\nabla\cdot\left(\frac{c}{V_A^2}\nabla_\perp\delta\phi\right)
+\mathbf{B_0}\cdot\nabla\left[\frac{1}{B_0^2}\nabla\cdot\left(B_0\nabla_\perp \delta A_{||}\right)\right]
-8\pi{\nabla\delta P\cdot\frac{\mathbf{b_0}\times\boldsymbol{\kappa}}{B_0}}
-\frac{4\pi}{c}\mathbf{\delta B}\cdot\nabla\left(\frac{J_{||0}}{B_0}\right)=0
\end{split}.
\end{flalign}
Next, by taking $\nabla_\perp\times\left[\boldsymbol{B_0}\times\cdots\right]$ on Eq. \eqref{momu_eq}, we have the dynamic equation for $\delta B_{||}$
\begin{flalign}\label{mhd-bpara}
\begin{split}
\frac{1}{V_A^2}\frac{\partial ^2\delta B_{||}}{\partial t^2}= \frac{4\pi}{B_0}\nabla_\perp^2\delta P +\nabla_\perp^2\delta B_{||},
\end{split}
\end{flalign}
where the inertia term on the LHS of Eq. \eqref{mhd-bpara} contributes to the high frequency CAWs and is often omitted in the low frequency regime $\omega\ll\Omega_{ci}$. It should also be pointed out that the higher order terms in $k_{||}/k_\perp$ are omitted in Eqs. \eqref{mhd-vorticity} and \eqref{mhd-bpara}.
The plasma parallel momentum equation remains the same with classic ideal full-MHD model, which is derived by taking the dot product between Eq. \eqref{momu_eq} and the unit vector of magnetic field $\mathbf{b_0}$
\begin{flalign}\label{mhd-sound}
\begin{split}
\rho_0\frac{\partial \delta u_{||}}{\partial t}
=-\mathbf{b_0}\cdot\nabla\delta P  - \frac{\mathbf{\delta B}}{B_0}\cdot\nabla P_0
\end{split},
\end{flalign}
where we also use the equilibrium force balance $\nabla P_0= \frac{1}{c}\mathbf{J_0}\times\mathbf{B_0}$ to derive above form .
Substituting Eq. \eqref{v_ideal_MHD} into Eq. \eqref{deltaP}, the MHD pressure equation in terms of potentials can be expressed as
\begin{flalign}\label{mhd-deltaP}
\frac{\partial \delta P}{\partial t} =
-\frac{c\mathbf{b_0}\times\nabla\delta\phi}{B_0}\cdot\nabla P_0
- 2\Gamma P_0c\nabla\delta\phi\cdot\frac{\mathbf{b_0}\times\boldsymbol{\kappa}}{B_0}
+\Gamma P_0 \frac{1}{B_0} \frac{\partial\delta B_{||}}{\partial t}
- \Gamma P_0\mathbf{B_0}\cdot\nabla\left(\frac{\delta u_{||}}{B_0}\right),
\end{flalign}
where we have dropped the term $-\frac{1}{B_0}\mathbf{b_0}\times\frac{\partial\mathbf{\delta A_\perp}}{\partial t}\cdot\nabla P_0$ on the RHS of Eq. \eqref{mhd-deltaP}, since the convection is mainly caused by the slow modes characterized with $|c\nabla_\perp\delta\phi|\gg|\partial_t\mathbf{\delta A_\perp}|$ in strongly magnetized plasmas. Eqs. \eqref{mhd-ohm}, \eqref{mhd-vorticity}-\eqref{mhd-deltaP} consist of a closed equation set as an ideal full-MHD framework in terms of potentials $\left(\delta\phi,\delta A_{||}, \delta B_{||}\right)$. Note that Eq. \eqref{mhd-bpara} reduces to the perpendicular force balance relation $4\pi\delta P + \delta B_{||}B_0 =0$ by removing the undesired fast wave (i.e., inertia term), which has been considered in the derivation of the interchange term in Eq. \eqref{mhd-vorticity} \cite{fu2006} and gives the leading $\delta B_{||}$ effect implicitly, i.e., drift reversal cancellation \cite{joiner2010, dong2017}. Thus Eq. \eqref{mhd-bpara} is redundant and can be dropped, meanwhile the $\delta B_{||}$ term in Eq. \eqref{mhd-deltaP} is $O\left(\beta\right)$ order small and also dropped, then we can arrive at the $\left(\delta\phi,\delta A_{||}\right)$ form in section \ref{2.2}.

Next we show a simple benchmark of above potential form of MHD equation set against the analytic dispersion relation in uniform plasma limit. By removing the equilibrium non-uniformities, the MHD shear Alfv\'en wave can be derived using Eqs. \eqref{mhd-ohm} and \eqref{mhd-vorticity} as
\begin{flalign}\label{saw}
	\begin{split}
		\omega^2 = k_{||}^2V_A^2
	\end{split},
\end{flalign}
 and the MHD slow and fast wave branches can be obtained from Eqs. \eqref{mhd-bpara}-\eqref{mhd-deltaP} as
 \begin{flalign}\label{slow_fast}
 	\omega^2 = \frac{k^2C_s^2 + k_\perp^2 V_A^2}{2}\left[1\pm\sqrt{1-\frac{4k_\perp^2k_{||}^2C_s^2V_A^2}{\left(k^2C_s^2+k_\perp^2V_A^2\right)^2}}\right],
 \end{flalign}
where $C_s = \sqrt{\Gamma P_0/\rho_0}$. Comparing Eqs. \eqref{saw} and \eqref{slow_fast} with the results from classic MHD model using $\left(\mathbf{E},\mathbf{B}\right)$ fields \cite{freidberg}, it is seen that the analytic dispersion relations of potential form are valid in the regime of $k_{||}\ll k_\perp$, which is consistent with the assumption made for the derivations of model equations of potential form. 

Hence, the MHD equations in terms of potentials $\left(\delta\phi,\delta A_{||}, \delta B_{||}\right)$, Eqs. \eqref{mhd-ohm}, \eqref{mhd-vorticity}-\eqref{mhd-deltaP}, can be used to compare with the comprehensive Landau-fluid model or other advanced models using potentials straightforwardly for validating simulation results and delineating the new physics effects. 

\section{Plasma dispersion relation using drift-kinetic model}\label{A4}
Here, we briefly explain the derivation of the drift-kinetic dispersion relation Eq. \eqref{DR-dk}. In uniform plasmas and magnetic field, the drift-kinetic equation for each particle species is
\begin{flalign}\label{dk}
	\begin{split}
		\left(\frac{\partial}{\partial t} + v_{||}\mathbf{b_0}\cdot\nabla\right) \delta f_\alpha
		=  \frac{Z_\alpha}{m_\alpha}\left(\mathbf{b_0}\cdot\nabla\delta\phi + \frac{1}{c}\frac{\partial\delta A_{||}}{\partial t}\right)\frac{\partial f_{\alpha0}}{\partial v_{||}}
	\end{split},
\end{flalign}
where the subscript $\alpha = i,e$ stands for the ion and electron species, $m_\alpha$, $Z_\alpha$ and $v_{||}$ represent particle mass, charge and parallel velocity, respectively. Considering $\delta n_\alpha = \int\mathbf{dv}\delta f_\alpha$ and $n_{\alpha0}\delta u_{||\alpha} = \int\mathbf{dv}v_{||}\delta f_\alpha$, Eqs. \eqref{poisson}, \eqref{Apara} and \eqref{dk} form a closed linear system of drift kinetic model in uniform plasmas. Applying the Fourier transforms $\partial_t\to -i\omega$ and $\mathbf{b_0}\cdot\nabla\to ik_{||}$, and substituting the Maxwellian equilibrium distribution $f_{\alpha0} = n_{\alpha0}\left(\frac{m_\alpha}{2\pi T_{\alpha0}}\right)^{3/2}exp\left(-\frac{m_\alpha v_{||}^2 + 2\mu B_0}{2T_{\alpha0}}\right)$ into Eq. \eqref{dk}, the perturbed distribution $\delta f_\alpha$ can be solved as
\begin{flalign}\label{delta_f}
	\begin{split}
		\delta f_\alpha = \frac{Z_\alpha}{T_{\alpha0}} \left(\frac{k_{||}v_{||}}{\omega- k_{||}v_{||}}\right)\left(\delta\phi - \frac{1}{c}\frac{\omega}{k_{||}}\delta A_{||}\right)f_{\alpha 0 }
	\end{split},
\end{flalign}
then the corresponding $\delta n_\alpha$ and $\delta u_{||\alpha}$ are calculated by integrating Eq. \eqref{delta_f} over the velocity space
\begin{flalign}\label{dn_alpha}
	\begin{split}
		\delta n_\alpha = -\frac{Z_\alpha n_{\alpha 0 }}{T_{\alpha0}}\left(\delta\phi - \frac{1}{c}\frac{\omega}{k_{||}}\delta A_{||}\right)\left[1 + \xi_\alpha Z\left(\xi_\alpha\right)\right]
	\end{split}
\end{flalign}
and
\begin{flalign}\label{du_alpha}
	\begin{split}
		\delta u_{||\alpha} = -\frac{Z_\alpha}{T_{\alpha0}}\frac{\omega}{k_{||}}\left(\delta\phi - \frac{1}{c}\frac{\omega}{k_{||}}\delta A_{||}\right)\left[1 + \xi_\alpha Z\left(\xi_\alpha\right)\right]
	\end{split},
\end{flalign}
where $\xi_\alpha = \omega/\left(\sqrt{2}k_{||}v_{th\alpha}\right)$, $v_{th\alpha} = \sqrt{T_{\alpha0}/ m_\alpha}$ is the thermal speed, and $Z\left(\xi_\alpha\right) = \frac{1}{\sqrt{\pi}}\int_{-\infty}^{+\infty}\frac{exp(-t^2)}{t-\xi_\alpha}dt$ is the plasma dispersion function. Substituting Eqs. \eqref{dn_alpha} and \eqref{du_alpha} into Eqs. \eqref{poisson} and \eqref{Apara} and after some algebra, we can arrive at the desired equation for the linear dispersion relation of drift-kinetic model, i.e.,
\begin{flalign}\label{DR-dk2}
	\begin{split}
		\left[\frac{\omega^2}{k_{||}^2V_A^2}-1\right]\left[1 + \xi_e Z\left(\xi_e\right)+\frac{T_{e0}}{T_{i0}}\frac{Z_i^2n_{i0}}{e^2n_{e0}}\left(1 + \xi_i Z\left(\xi_i\right)\right)\right] 
		= \frac{Z_i^2n_{i0}}{e^2n_{e0}}k_\perp^2\rho_s^2
	\end{split}.
\end{flalign}
It is seen that the SAW and ISW are coupled through finite $k_\perp^2\rho_s^2$ and decoupled in the long wavelength limit of $k_\perp^2\rho_s^2\to 0$.

\end{document}